\documentclass[journal = jpccck]{achemso}
\usepackage{graphics,graphicx}
\usepackage{tabularx}
\usepackage[T1]{fontenc}
\usepackage{cancel}
\usepackage{times}
\usepackage{amssymb,amsfonts,amsmath}

\newlength{\figurewidth}
\usepackage[version=4]{mhchem}

\newlength{\figwidthcol}
\newlength{\figwidthtwocol}
\usepackage{color}

\title{Structural Diversity and Superconductivity in S-P-H Ternary Hydrides Under Pressure}

\author{Nisha Geng}
\affiliation{Department of Chemistry, State University of New York at Buffalo, Buffalo, NY 14260-3000, USA}
\author{Tiange Bi}
\affiliation{Department of Chemistry, State University of New York at Buffalo, Buffalo, NY 14260-3000, USA}
\author{Eva Zurek}
\email{ezurek@buffalo.edu}
\affiliation{Department of Chemistry, State University of New York at Buffalo, Buffalo, NY 14260-3000, USA}

\begin{document}
\begin{abstract}
Evolutionary structure searches revealed a plethora of stable and low-enthalpy metastable phases in the S-P-H ternary phase diagram under pressure. A wide variety of crystalline structure types were uncovered ranging from those possessing one-dimensional chains, two-dimensional sheets based on S-H or S-P-H square lattices as well as S-H or P-H honeycombs, and cage-like structures. Some of the cage-like structures could be derived from doping the high-pressure high-temperature superconducting $Im\bar{3}m$ H$_3$S phase with phosphorous. Most of the discovered compounds were metallic, however those derived from $Im\bar{3}m$ H$_3$S lattices with low levels of P-doping were predicted to possess the highest superconducting critical temperatures ($T_c$s). The propensity for phosphorous to assume octahedral coordination, as well as the similar radii of sulfur and phosphorous are key to maintaining a high density of states at the Fermi level in $Im\bar{3}m$ S$_{0.875}$P$_{0.125}$H$_3$, whose $T_c$ was estimated to be similar to that of H$_3$S at 200~GPa.
\end{abstract}

\maketitle

\newpage

\section{Introduction}
The recent successes in the search for materials with ever-higher superconducting critical temperatures ($T_c$s) can be traced back to ideas planted by the late Neil Ashcroft. Building on his earlier prediction of phonon-mediated high-temperature superconductivity in metallic hydrogen \cite{Ashcroft:1968a}, in 2004 Ashcroft elegantly showed how chemical doping may be employed to decrease the pressure required to reach a metallic and superconducting state \cite{Ashcroft:2004a}. This proposal ultimately led to the experimental discovery of superconductivity in H$_3$S \cite{Drozdov:2015}, LaH$_{10}$ \cite{Somayazulu:2019,Drozdov:2019}, YH$_6$ \cite{Troyan-YH4,Kong:arxiv}, YH$_9$ \cite{Kong:arxiv,Snider:2021} and room temperature superconductivity in an unknown C-S-H phase \cite{Snider:2020}. 

Because of the experimental difficulties inherent in high pressure research, theoretical calculations have been integral in suggesting targets for synthesis and for interpreting experimental measurements \cite{Zurek:2014i,Zurek:2019k}. Advances in \emph{ab initio} crystal structure prediction techniques and computational methods to approximate $T_c$ have enabled the \emph{in silico} search for superconductivity in high pressure hydrides \cite{Livas:2020-Review}. By now, the phase diagrams of most binary hydrides have been explored computationally \cite{Zurek:2018d,Wang:2018,Zurek:2019h}. Ternary hydrides are the next frontier, however the combinatorial possibilities, potential for large unit cells, and role of metastability results in many challenges \cite{Boeri:2021,Zurek:2021k}.

One way that new ternary superconductors could potentially be made is by combining two binary hydrides that are thought to be good superconductors.  For example, the extraordinary $T_c$ values predicted for CaH$_6$ \cite{Wang:2012} and measured for SH$_3$ \cite{Drozdov:2015,Duan:2014} inspired the computational search for superconductivity in Ca$_x$S$_y$H$_z$ phases \cite{Yan:2020}.  A $P\bar{6}m2$ CaSH$_3$ compound built from CaH$_2$ layers separated by honeycomb H-S layers was predicted to have a $T_c$ reaching 100~K at 128~GPa. Inspired by the prediction of high $T_c$ in metal superhydrides with clathrate cages \cite{Peng:2017h,Liu:2017}, the (La/Y)H$_{10}$ and (La/Y)H$_6$ ternaries were synthesized, and their maximum measured $T_c$s were 253 and 237~K, respectively \cite{Semenok:2021b}. Another strategy that could be used to find promising ternary superconductors is to begin with a binary phase known to have a large $T_c$ and dope it with a third (or fourth) element.  The doping could potentially increase $T_c$ (c.f.\ Li$_2$MgH$_{16}$ \cite{Sun:2019} versus MgH$_{16}$ \cite{Zurek:2012m}) or extend the region of phase stability to lower pressures (c.f.\ LaBH$_8$ \cite{diCataldo:2021,Liang:2021} or LaBeH$_8$ \cite{Zhang:2021} versus LaH$_{10}$ \cite{Liu:2017,Peng:2017h}).  

It has been proposed that doping can be employed to tune the position of the Fermi level ($E_F$) so that it lies on a maximum in the density of states (DOS) thereby increasing the electron phonon coupling parameter, $\lambda$, and concomitantly the $T_c$. This mechanism ought to be particularly useful for enhancing the $T_c$ in H$_3$S, whose $E_F$ is just shy of the top of a peak caused by the presence of two van Hove singularities (vHs) that straddle $E_F$. Computations using the virtual crystal approximation (VCA), which creates alchemical atoms whose properties vary smoothly upon doping, have explored the effect of doping on the $T_c$ of  H$_3$S. Heil and Boeri found that $T_c$ could be enhanced by replacing sulfur by more electronegative elements such as oxygen or  halogen atoms \cite{Heil:2015a}, whereas Ge \emph{et al.} and Hu \emph{et al.} concluded that $T_c$s above the freezing point of water could be obtained by low dopings of phosphorous \cite{Ge:2016}, silicon \cite{Ge:2016,Ge:2020a}, and most recently carbon \cite{Ge:2020a,Hu:2020}. However, other VCA-based studies found that partial substitution of sulfur with another $p$-block element did not enhance $T_c$ significantly, or not at all \cite{Fan:2016,Wang:2021a}.

The VCA model is advantageous since it enables the study of low doping levels that can only be modelled by large unit cells whose $T_c$s cannot be computed from first principles because of the immense expense. At the same time, the VCA assumes the band structure will not be perturbed by doping, and only the position of $E_F$ will change (rigid band model). Recent theoretical studies have called this assumption into question. For example, calculations on C$_x$S$_{1-x}$H$_3$ and C$_x$S$_{1-x}$H$_{3+x}$ supercells with doping levels ranging from 1.85-25\% illustrated that incorporation of carbon into the H-S lattice decreases the number of bands at $E_F$ by breaking degeneracies and localizing electrons in covalent C-H bonds \cite{Zurek:2021i}. Calculations on phases where half of the SH$_3$ units were substituted by methane resulting in a CSH$_7$ stoichiometry, yielded maximal $T_c$s of 181~K at 100~GPa~\cite{Sun:2020}and 194~K at 150~GPa~\cite{Cui:2020}. Theory has shown that Se \cite{Amsler:2019,Liu:2018prb} and Cl \cite{nakanishi:2018,Guan:2021a} dopings do not raise the superconducting critical temperature of H$_3$S. And, phosphorous dopings have been shown to both increase and decrease its $T_c$. For example, a 6.25\% phosphorus doping resulted in a $T_c$ of 212~K \cite{nakanishi:2018} or 262~K at 200~GPa \cite{Guan:2021a}, a 12.5\% phosphorus doped structure was predicted to possess a $T_c$ of 194~K at 200~GPa \cite{nakanishi:2018}  or 190~K at 155~GPa \cite{Durajski:2018}, and a 50.0\% phosphorus doped structure yielded a $T_c$ of 89~K at 200~GPa \cite{Tsuppayakorn:2021} or 157~K at 155~GPa \cite{Durajski:2018}. Another drawback of both the VCA \cite{Fan:2016,Ge:2016} and the supercell models with S$\rightarrow$P substitutions \cite{nakanishi:2018,Durajski:2018,Guan:2021a} is that they did not explore the potential energy landscape for the compositions considered, and the S-P-H ratios were fixed. Moreover, the studies carried out with both methods employed structures derived from $Im\overline{3}m$ or $R3m$ H$_3$S cells, which may not form in experiment, especially if other novel S-P-H ternary phases are thermodynamically preferred \cite{Livas:2020-Review}.

To overcome these limitations, herein we perform a thorough theoretical study that employs crystal structure prediction techniques to identify the most stable ternary phases for a wide S$_x$P$_y$H$_z$ composition range. The electronic structure including the density of states, electron localization functions, electronic and phonon band structures are analyzed. Four different families of structures including those comprised of cages, two dimensional sheets with hexagonal or other motifs, and one-dimensional molecular chains are found. Unlike our previous results on carbon doping of H$_3$S, here we find at least one system where phosphorous doping does not decrease the DOS at $E_F$. This $Im\bar{3}m$ S$_7$PH$_{24}$ compound is estimated to have a $T_c$ comparable to its parent $Im\bar{3}m$ H$_3$S structure at 200~GPa. We hope our study can guide the experimental discovery and characterization of new ternary superconducting S-P-H compounds under pressure.

\section{Computational Details}
Crystal structure prediction (CSP) searches were performed to find stable and low enthalpy metastable phases with  SPH$_n$~($n=1-6$), S$_{x}$PH and SP$_{y}$H ($x,y=2,3$) stochiometries, as well as those corresponding to phosphorus doped H$_3$S phases at 100, 150 and 200~GPa using 1-4 formula units in the simulation cell. The stoichiometries of the doped phases we considered were S$_x$P$_{1-x}$H$_3$ ($x=0.125, 0.2, 0.25, 0.33, 0.4, 0.6, 0.67, 0.75, 0.8$), which were modelled by S$_7$PH$_{24}$, S$_4$PH$_{15}$, S$_3$PH$_{12}$, S$_2$PH$_9$, S$_3$P$_2$H$_{15}$, S$_2$P$_3$H$_{15}$, SP$_2$H$_{9}$, SP$_3$H$_{12}$, and SP$_4$H$_{15}$ cells, respectively. These were chosen based on a similar study for S$_x$Se$_{1-x}$H$_3$ systems\cite{Amsler:2019}. In this case the EA searches were seeded with doped supercells from the previously reported $Im\overline{3}m$ H$_3$S and $R3m$ H$_3$S  structures\cite{Duan:2014}. The CSP searches were performed using the open-source evolutionary algorithm (EA)  \textsc{XtalOpt} \cite{Zurek:2011a,xtalopt-web,Zurek:2020i} version 12 \cite{Zurek:2018j}. The initial generation consisted of random symmetric structures that were created by the  \textsc{RandSpg} algorithm \cite{Zurek:2016h}, and the minimum interatomic distance between S-S, S-H, S-P, H-H, H-P, and P-P atoms were constrained to 0.88, 0.71, 0.93, 0.53, 0.76 and 0.98 \AA{} using a uniform scaling factor of 0.5 multiplied by tabulated covalent radii. Duplicate structures were identified via the  \textsc{XtalComp} algorithm \cite{Zurek:2011i} and discarded from the breeding pool. The lowest enthalpy structures in each EA run were relaxed from 100 to 200~GPa and their relative enthalpies and equations of states are given in the Supplementary Information Figs.\ S1-6. 

Geometry optimizations and electronic structure calculations including band structures, the densities of states (DOS), electron localization functions (ELFs), and Bader charges were carried out using density functional theory (DFT) as implemented in the Vienna \textit{ab-initio} Simulation Package (VASP) version 5.4.1 \cite{Kresse:1993a, Kresse:1999a}, with the gradient-corrected exchange and correlation functional of Perdew{-}Burke{-}Ernzerhof (PBE) \cite{Perdew:1996a}. The crystal orbital Hamilton population (COHP) and the negative of the COHP integrated to the Fermi level (-iCOHP) was calculated using the LOBSTER package to analyze the bonding of selected phases. \cite{Maintz:2016} 

The projector augmented wave (PAW) method \cite{Blochl:1994a} was employed, and the S 3s$^2$3p$^4$, P 3s$^2$3p$^3$ and H 1s$^1$ electrons were treated explicitly in all of the calculations. The plane-wave basis set energy cutoffs were 350-400 eV in the EA searches and 700 eV in precise geometry optimizations. The $k$-point meshes were generated using the $\Gamma$-centered Monkhorst-Pack scheme, while the number of divisions along each reciprocal lattice vector was selected so that the product of this number with the real lattice constant was 30~\AA{} in the EA searches and 50~\AA{} for precise geometry optimizations.

Phonon calculations were carried out via the supercell approach \cite{Parlinski:1997,Chaput:2011} using the VASP package coupled to the PHONOPY code\cite{Togo:2015}. The electron-phonon coupling (EPC) calculations were carried out using the Quantum Espresso (QE) program \cite{Giannozzi:2009}. The S 3s$^2$3p$^4$, P 3s$^2$3p$^3$ and H 1s$^1$ pseudopotentials were obtained from the PSlibrary \cite{DalCorso:2014}, and generated by the method of Trouiller-Martins \cite{Troullier:1991}. The plane-wave basis set energy cutoffs were chosen to be 60~Ry. The Brillouin zone sampling scheme of Methfessel-Paxton \cite{Methfessel:1989} was applied. The EPC parameter, $\lambda$, was calculated using a set of Gaussian broadenings with an increment of 0.005~Ry from 0.0 to 0.500~Ry. The broadening for which $\lambda$ was converged to within 0.05~Ry is provided in Table S5, along with the $k$ and $q$ grids that were deemed to be converged. The $T_c$ was estimated using the Allen-Dynes modified McMillan equation \cite{Allen:1975} with a renormalized Coulomb potential, $\mu^*$, of 0.1. For phases with $\lambda$ greater than 1.3 more accurate estimates of the $T_c$ were also calculated by numerically solving the Eliashberg equations. \cite{Eliashberg:1960}

\section{Results and Discussion}
\begin{figure*}
		\includegraphics[width=\textwidth]{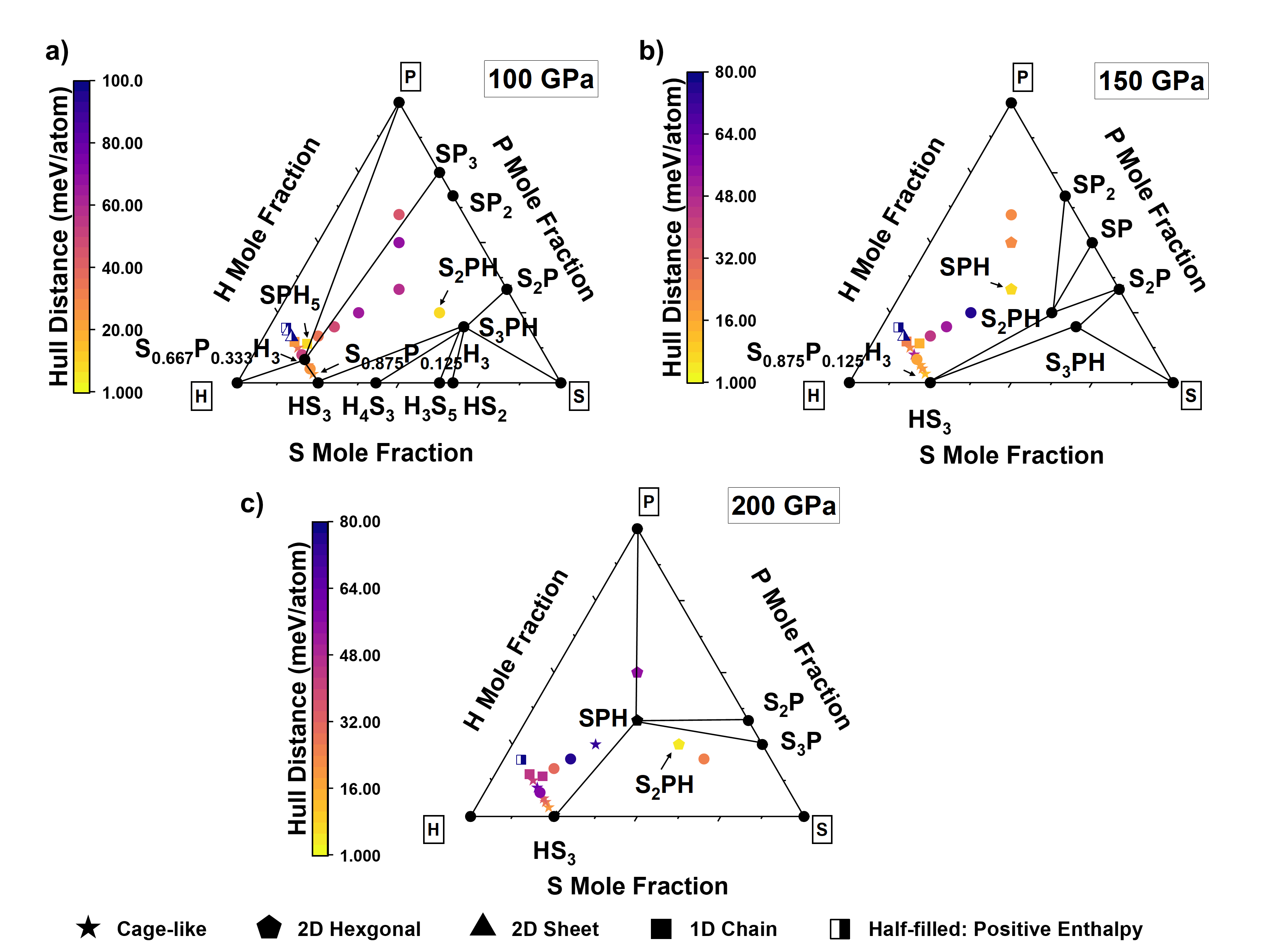}
	\caption{Distance from the convex hull of the S-P-H phases considered 
	with respect to elemental S, P, and \textonehalf~H$_2$ as computed with the PBE functional at 100, 150, and 200~GPa. The thermodynamically stable phases are shown in black while the metastable phases are represented by different colors based on their distance from the hull. The phases with positive enthalpies are shown with a hall-filled shape. The phases containing special geometrical motifs such as cage-like structures, 2D hexagonal or other sheets, and 1D chains are labelled as stars, pentagons, triangles, and squares, respectively. $\Delta H_F$ was calculated using the enthalpies of the experimentally or theoretically known structures: simple cubic (sc, 100~GPa)\cite{Jamieson:1963,Kikegawa:1983,Akahama:1999} and simple hexagonal (sh, 150--200~GPa) \cite{Akahama:1999} for P; $P6_3/m$ (100~GPa)\cite{Pickard:2007s} and $C2/c$ (150--200~GPa) \cite{Pickard:2007s} for H; $\beta$-Po \cite{Luo:1993} for S (100--200~GPa) since the crystal structure of S above 83~GPa is still disputed \cite{Zakharov:1995,Gavryushkin:2017a,Kokail:2016}; $C2/c$ (200~GPa) for S$_3$P\cite{Liu:2019}; $C2/m$ (100--150~GPa) and $P\overline{3}m1$ (200~GPa) for S$_2$P\cite{Liu:2019}; $P2_1/m$ (150~GPa) for SP\cite{Liu:2019}; $C2/m$-I (100~GPa) and $C2/m$-II (150~GPa) for SP$_2$\cite{Liu:2019}; $C2/m$ (100~GPa) for SP$_3$ \cite{Liu:2019}; $C2/c$ (100~GPa), $R3m$ (150~GPa), and $Im\overline{3}m$ (200~GPa) for H$_3$S \cite{Drozdov:2015,Duan:2014,Li:2016-S}; $Pbcm$ (100~GPa) and $C2c$ (150--200~GPa) for HS$_2$ \cite{Kruglov:2017,Errea:2015}; $Pnma$ (100~GPa) for H$_4$S$_3$ \cite{Li:2016-S}; $C2/m$ (100~GPa) for H$_3$S$_5$ \cite{Goncharov:2016}.
		\label{fig:SPH-convex}}
\end{figure*}

The enthalpies of formation, $\Delta H_F$, of the most stable S-P-H ternary phases found in our EA searches for a given stoichiometry at 100, 150 and 200~GPa were calculated and employed to construct the convex hull.  If the  $\Delta H_F$ of a structure falls on the convex hull, it is a thermodynamically stable phase. However, previous theoretical and experimental work on binary hydrides has illustrated that synthesizability is not only determined by the $\Delta H_F$ estimated within the static lattice approximation. It turns out that the synthesis pathway, which includes the choice of the precursors and annealing schemes, is also key in determining the products that are formed. For example, despite the fact that Ca$_2$H$_5$ lies 20~meV/atom above the convex hull, this binary hydride has been synthesized \cite{Zurek:2018b}. Moreover, even though DFT calculations predict that hydrides of phosphorous are not thermodynamically stable under pressure \cite{Zurek:2017c,Shamp:2016,Flores:2016}, a superconducting PH$_n$ phase \cite{Drozdov:2015b} and a van der Waals (PH$_3$)$_2$H$_2$ crystalline compound \cite{Ceppatelli} have been synthesized. Moreover, a number of H$_x$S$_y$ superconductors have been made  \cite{Yao-S-review:2018}, and various stoichiometries including H$_3$S \cite{Duan:2014,Verma:2018a},  H$_2$S \cite{Li:2014,Akashi:2015-S}, HS$_2$ \cite{Errea:2015}, H$_4$S$_3$ \cite{Li:2016-S}, H$_5$S$_2$ \cite{Ishikawa:2016}, exotic Magn\'{e}li phases \cite{Akashi:2016a}, and self-ionized (SH$^-$)(H$_3$S)$^+$ perovskites \cite{Gordon:2016} that may have long modulations \cite{Majumdar:S-2017,Majumdar:S-2019} have been proposed. Because metastable binary S-H and P-H phases have been synthesized, herein we have chosen to analyze structures that fall on the convex hull (Fig.\ \ref{fig:SPH-convex}), as well as several metastable phases with unique structural characteristics that are within the realm of synthesizability.

At 100~GPa both S$_2$PH$_9$, which can be considered a P-doped H$_3$S phase with a S$_{0.67}$P$_{0.33}$H$_3$ composition, and S$_3$PH lie on the hull. In addition to S$_3$PH, S$_2$PH also comprises the 150~GPa hull, whereas only SPH falls on the hull at 200~GPa. Noteably, most of the compositions studied were within the 70~meV/atom threshold corresponding to the 90$^\textrm{th}$ percentile of DFT-calculated metastability for inorganic crystalline materials at 1~atm \cite{materialsproject}, but the $\Delta H_F$ for three phosphorus-rich compounds,  SP$_2$H$_9$ (S$_{0.33}$P$_{0.67}$H$_3$), SP$_3$H$_{12}$ (S$_{0.25}$P$_{0.75}$H$_3$) and SP$_4$H$_{15}$	(S$_{0.2}$P$_{0.8}$H$_3$), were positive with respect to the elemental phases at all pressures. SP$_3$H was also unstable with respect to H$_2$, P and S at 200~GPa. The structural parameters and the distance from the convex hull for all S-P-H ternary phases considered are provided in Table S6-8 and S2, respectively. 

Because the enthalpy of formation of the binary hydrides of phosphorous is positive above 40~GPa \cite{Zurek:2017c,Shamp:2016,Flores:2016}, none of the previously predicted phases are plotted on Fig.\ \ref{fig:SPH-convex}. For the hydrides of sulfur the $C2/c$ H$_3$S\cite{Li:2016-S}, $Pnma$ H$_4$S$_3$ \cite{Li:2016-S}, $Pbcm$ HS$_2$\cite{Kruglov:2017} and $C2/m$ H$_3$S$_5$\cite{Goncharov:2016} phases were employed to construct the 100~GPa hull, while $R3m$ and $Im\overline{3}m$ H$_3$S \cite{Duan:2014} were used to build the 150 and 200~GPa hulls, respectively. For the P-S phases we employed structures previously identified via crystal structure prediction~\cite{Liu:2019}: $C2/m$ SP$_3$, $C2/m$-I SP$_2$ and $C2/m$ S$_2$P lay on the 100~GPa hull; $C2/m$-II SP$_2$, $P2_1/m$ SP and $C2/m$ S$_2$P lay on the hull at 150~GPa; $P\overline{3}m1$ S$_2$P and $C2/c$ S$_3$P lay on the hull at 200~GPa. 

In our work we found stable and metastable ternary S-P-H species with a variety of distinct structural motifs including: (i) cage-like geometries, e.g.\ S$_7$PH$_{24}$	(S$_{0.875}$P$_{0.125}$H$_3$), S$_4$PH$_{15}$	(S$_{0.8}$P$_{0.2}$H$_3$), S$_3$PH$_{12}$	(S$_{0.75}$P$_{0.25}$H$_3$), S$_3$P$_2$H$_{15}$	(S$_{0.6}$P$_{0.4}$H$_3$), SPH$_6$ (S$_{0.5}$P$_{0.5}$H$_3$) and SPH$_2$; (ii) two-dimensional hexagonal sheets, e.g.\ S$_2$PH and SP$_2$H; (iii) other types of two-dimensional sheets, e.g.\ $P\overline{1}$ SP$_2$H$_{9}$	(S$_{0.33}$P$_{0.67}$H$_3$) and SP$_3$H$_{12}$	(S$_{0.25}$P$_{0.75}$H$_3$);  and (iv) one-dimensional chains, e.g.\ SPH$_5$, S$_2$P$_3$H$_{15}$	(S$_{0.4}$P$_{0.6}$H$_3$) and SP$_4$H$_{15}$	(S$_{0.2}$P$_{0.8}$H$_3$). We also studied other thermodynamically stable phases (e.g.\ S$_3$PH and S$_2$PH$_{9}$	(S$_{0.67}$P$_{0.33}$H$_3$)), and other superconducting phases (e.g.\ SPH$_3$ and SPH$_4$). The structural peculiarities of many of these are described in the following sections to showcase the rich potential energy landscape and geometrical complexity that is possible. However, for brevity we do not discuss dynamically stable phases that either did not possess any distinct structural motifs, or those that are only slight variations of presented phases. Information on some of these, e.g. SPH$_n$ $(n=3-5)$ and SP$_4$H$_{15}$ (S$_{0.2}$P$_{0.8}$H$_3$), can be found in Fig.\ S42.

\begin{table*}[ht!]
	\caption{$\Delta H$ (meV/atom) for the reaction $\text{H}_3\text{S}+x\text{P}\rightarrow \text{H}_3\text{S}_{1-x}\text{P}_x+x\text{S}$ for different doping levels, $x$, at various pressures.} 
	\centering % center the table
\begin{tabular}{lcccccccccc}
    \hline\hline
$P$/$x$  & 12.5\%${^a}$ & 20\%${^a}$ & 25\%${^a}$ & 33\%${^b}$ & 40\%${^a}$ & 50\%${^a}$ & 60\%${^c}$ & 67\%${^c}$ & 75\%${^c}$ & 80\%${^d}$ \\
    \hline
100      & 34.3 (23.9$^e$)    & 41.5   & 87.4   & 48.1   & N/A   & 82.1   & 133.0 & 119.4  & 139.1  & 79.8   \\
150      & 26.6  (31.2$^e$)   & 43.4   & 96.4   & 47.4   & 57.2   & 93.7   & 136.3  & 120.2  & 147.0  & 79.5   \\
200      & 40.0  (36.5$^e$/37.0$^f$)  & 55.3   & 116.2  & 66.1   & 100.7  & 123.1  & 149.7  & 142.7  & 175.6  & 107.1 \\
    \hline
    \end{tabular} \\
    $^a$ Illustrated in Fig.\ \ref{fig:structure-cage}\\
    $^b$ Illustrated in Fig.\ \ref{fig:hullstructures}\\
    $^c$ Illustrated in Fig.\ \ref{fig:structure-2Dsheets}\\
    $^d$ Illustrated in Fig. S42 \\
    $^e$ Ref.\ \cite{nakanishi:2018} \\
        $^f$ Ref.\ \cite{Guan:2021a} \\
    \label{tab:H-SPH}
\end{table*} 

To guage the stability of the P doped phases, we considered the enthalpy of the reaction $\text{H}_3\text{S}+x\text{P}\rightarrow \text{H}_3\text{S}_{1-x}\text{P}_x+x\text{S}$ (Table \ref{tab:H-SPH}). Our results for S$_{0.875}$P$_{0.125}$H$_3$ (S$_7$PH$_{24}$) were similar to those calculated for the same stoichiometry by Nakanishi and co-workers  \cite{nakanishi:2018}. The relatively small positive $\Delta H$ values suggest that this phase could potentially be synthesized, especially if configurational entropy is able to overcome the enthalpic penalty, as is the case for numerous multicomponent systems \cite{Toher:2019}. Generally speaking, as the doping level increases, the ternaries become progressively destabilized with the sulfur-rich and phosphorours-rich phases possessing $\Delta H$ values ranging from 26.6-116.2~meV/atom and 79.8-175.6~meV/atom, respectively.  S$_2$PH$_{9}$	(S$_{0.67}$P$_{0.33}$H$_3$), and S$_3$P$_2$H$_{15}$ (S$_{0.6}$P$_{0.4}$H$_3$), whose structural peculiarities will be discussed below, did not follow this trend. 

%The S$_7$PH$_{24}$	(S$_{0.875}$P$_{0.125}$H$_3$) found in our search at 150~GPa had a lower enthalpy of 4.6 meV/atom than Nakanishi et.al\cite{nakanishi:2018}. 

\subsection{Cage Like Structures}

\begin{figure*}
\begin{center}
\includegraphics[width=\textwidth]{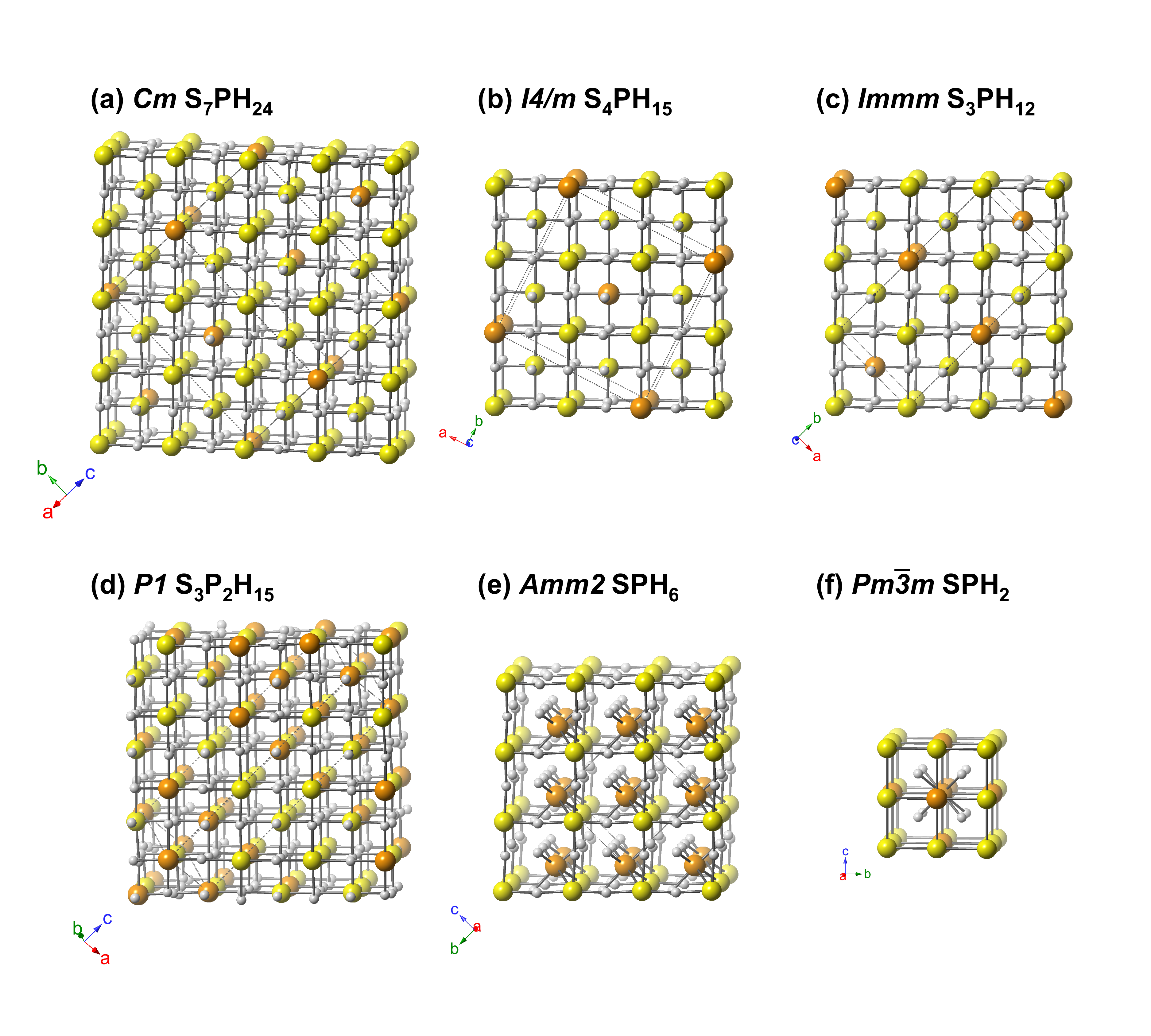}
\end{center}
\caption{Illustrations of predicted S-P-H phases with cage-like structures: (a) $Cm$ S$_7$PH$_{24}$ (S$_{0.875}$P$_{0.125}$H$_3$) at 150~GPa, (b) $I4/m$ S$_4$PH$_{15}$ at 200~GPa, (c) $Immm$ S$_3$PH$_{12}$ at 200~GPa , (d) $P1$ S$_3$P$_2$H$_{15}$ at 150~GPa, (e) $Amm2$ SPH$_6$ at 150~GPa, and (f) $Pm\overline{3}m$ SPH$_2$ at 200~GPa. S/P/H atoms are
colored yellow/orange/white.}  
\label{fig:structure-cage}
\end{figure*}

The discovery of record breaking $T_c$ in H$_3$S is a milestone in the search for room-temperature superconducting materials. Theoretical calculations predicted that $Im\overline{3}m$ H$_3$S is the most stable phase at 200~GPa with a $T_c$ of 191-204~K\cite{Duan:2014}, in-line with independent experiments that measured a $T_c$ of 203~K when hydrogen sulfide was compressed to 155~GPa~\cite{Drozdov:2015}. The S atoms in $Im\overline{3}m$ H$_3$S, which are separated symmetrically by hydrogen atoms, form a bcc lattice. Another structurally related H$_3$S phase with $R3m$ symmetry, which contains a pyramidal H$_3$S unit with a calculated $T_c$ of 155-166~K at 130~GPa, was also proposed \cite{Duan:2014}. Therefore, we seeded all of our S$_x$P$_{1-x}$H$_3$ EA searches with supercells of both $Im\overline{3}m$ and $R3m$ H$_3$S where some of the sulfur atoms were replaced by phosphorous atoms. Several cage-like structures including S$_7$PH$_{24}$ (S$_{0.875}$P$_{0.125}$H$_3$), S$_4$PH$_{15}$	(S$_{0.8}$P$_{0.2}$H$_3$), S$_3$PH$_{12}$	(S$_{0.75}$P$_{0.25}$H$_3$), S$_3$P$_2$H$_{15}$	(S$_{0.6}$P$_{0.4}$H$_3$), SPH$_6$ (S$_{0.5}$P$_{0.5}$H$_3$) and SPH$_2$, which resembled $Im\overline{3}m$ and $R3m$ H$_3$S, were found and are discussed below (see Fig. \ref{fig:structure-cage}).

A $Cm$ S$_7$PH$_{24}$ (S$_{0.875}$P$_{0.125}$H$_3$) structure with 1 forumula unit in the primitive cell was found via CSP, and it also possessed the lowest enthalpy at 100~GPa. This phase lay only 24.5 and 12.2~meV/atom above the convex hull at 100 and 150~GPa, respectively. It could be viewed as a distorted version of the $R3m$ cell considered by Nakanishi~\cite{nakanishi:2018}, and the distortion stabilized the structure by 14.9 and 7.0~meV/atom at 100 and 150~GPa, respectively. By 200~GPa the high-symmetry $Im\overline{3}m$ phase became preferred, and it lay 23.6~meV/atom above the convex hull. The $Cm$ phase contained SH$_3$ and SH$_4$ structural motifs, along with PH$_6$ units whose geometry deviated ever-so-slightly from ideal octahedral coordination.  In the higher pressure $Im\bar{3}m$ phase these motifs were forced close enough to each other so they formed multi-centered bonds where every S and P atom was octahedrally coordinated with SH-P and PH-S distances of 1.462 and 1.522~\AA{}, respectively, at 200~GPa. The HS-H distances in this phase were the same as within $Im\bar{3}m$ H$_3$S at the same pressure, 1.492~\AA{}. The Bader charge on the PH$_6$ unit within S$_7$PH$_{24}$ was calculated to be -0.09$e$ at 150~GPa and 0.03$e$ at 200~GPa.

\begin{figure}
\begin{center}
\includegraphics[width=1\columnwidth]{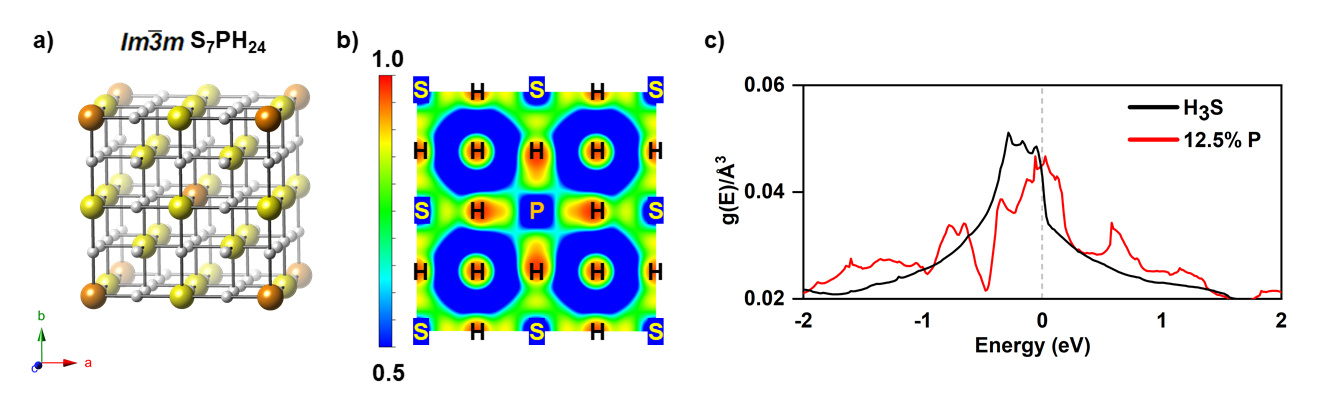}
\end{center}
\caption{(a) $Im\overline{3}m$ S$_7$PH$_{24}$ (S$_{0.875}$P$_{0.125}$H$_3$) at 150~GPa. (b) The Electron Localization Function (ELF) contour plotted in a plane that bisects the S-H-P contacts. (c) The density of states (DOS) plot near the Fermi ($E_F$) level of $Im\bar{3}m$ H$_3$S (black) and $Im\bar{3}m$ S$_7$PH$_{24}$ (red) at 200~GPa. The value of the DOS at $E_F$ is 0.043 and 0.045 states eV$^{-1}$/\AA{}$^3$ (0.063 and 0.067 states eV$^{-1}$/valence electron), respectively. }  
\label{fig:dos-elf}
\end{figure}

Our calculations show that the peak in the DOS of H$_3$S lies 0.28~eV below $E_F$ at 200~GPa. Assuming a rigid band model, this peak could be accessed by $\sim$12\% P doping, which corresponds well to the S$_{0.875}$P$_{0.125}$H$_3$ stoichiometry (Fig.\ \ref{fig:dos-elf}). In contrast to our results on H$_3$S with low levels of S$\rightarrow$C and SH$_3{\rightarrow}$CH$_4$ replacements \cite{Zurek:2021i}, the DOS at $E_F$ ($g(E)$/\AA{}$^3$) within $Im\bar{3}m$ S$_7$PH$_{24}$ is about the same as that of the undoped binary. A Bader analysis, which typically underestimates the formal charge, suggested a charge of +0.25 on CH$_6$, whereas PH$_6$ is nearly neutral. Moreover, a plot of the electron localization function (ELF) is indicative of the formation of multicentered P-H-S bonds, whereas the ELF plot for $O_h$-CS$_{15}$H$_{48}$ suggested the presence of strong C-H bonds, with the positively charged CH$_6$ motif weakly interacting with the nearby sulfur atoms. The different behavior of P versus C doping can be traced back to their elemental properties. The atomic radius of phosphorus (98~pm) is slightly larger than that of sulfur (87~pm), but the radius of carbon is markedly smaller (67~pm). Moreover, whereas the electronegativity of sulfur and carbon is about the same, phosphorous is somewhat less adept at attracting electrons. As a result, low levels of P-doping have a markedly different effect on the electronic structure of H$_3$S as compared to small amounts of C-doping. This finding suggests that choosing the right element for chemical substitution is key towards designing a material with an electronic structure that is particularly conducive towards superconductivity. 

To further interrogate the bonding in the P-doped structure the negative of the crystal orbital Hamilton population integrated to the Fermi level (-iCOHP) was calculated. In $Cm$ S$_7$PH$_{24}$ at 150~GPa five of the P-H bonds were stronger than the S-H bonds (4.18-4.23~eV/bond versus 1.97-2.12~eV/bond) and one bond was about the same strength (3.44 versus 3.13~eV/bond). Higher pressures promote multi-centered bonding so that in $Im\bar{3}m$ S$_7$PH$_{24}$ at 200~GPa the PH-S and SH-P -iCOHPs were almost equivalent (3.22 versus 3.74~eV/bond), and quite comparable to the bond strength within undoped $Im\bar{3}m$ H$_3$S at this pressure (3.67~eV/atom). Again, this differs markedly from our previous results, which showed that carbon doping decreased the CH-S bond strength to 2.00~eV/bond. Because P-doping does not perturb the structure nor the electronic structure of the H$_3$S lattice to the same degree as C-doping, it does not decrease the metallicity of H$_3$S significantly, or at all, and the electrons remain delocalized throughout the S-P-H lattice at 200~GPa. In contrast to our previous findings on S$\rightarrow$C substitutions in an octahedral coordination environment \cite{Zurek:2021i}, which decreased the $T_c$ from 174 to 130~K at 270~GPa, we would therefore expect the $T_c$ of $Im\bar{3}m$ S$_7$PH$_{24}$ to be closer to that of $Im\bar{3}m$ H$_3$S at 200~GPa, and perhaps even to surpass it. This hypothesis will be tested below.

S$_4$PH$_{15}$ (S$_{0.8}$P$_{0.2}$H$_3$) possesed the second highest S-concentration of the doped S$_{1-x}$P$_x$H$_3$ phases that were considered. $P1$, $I4$, and $I4/m$ symmetry structures were the most stable at  100, 150, and ~200 GPa, respectively, and they lay only 25.7, 20.2, and 29.0 meV/atom above the convex hull at these pressures. $P1$ S$_4$PH$_{15}$ contains four H$_2$, one PS$_2$H$_2$ and one (SH$_2$)$_2$H unit in the primitive cell. We do not show it in Fig.\ \ref{fig:structure-cage} since it is not a cage-like structure, and it is semiconducting at 100~GPa. The $I4$ and $I4/m$ symmetry phases are related to each other, and both can be considered as cage-like structures similar to $Cm$ S$_7$PH$_{24}$. At 150~GPa the H-S distances measured 1.415-1.667~\AA{}, shortening slightly to 1.412-1.590~\AA{} at 200~GPa (c.f.\ 1.492~\AA{} in $Im\overline{3}m$ H$_3$S at 200~GPa). The ideal octahedral geometry about the P atom was somewhat distorted in the $I4/m$ phase with four P-H bonds measuring 1.427~\AA{}, and the remaining two measuring 1.483~\AA{}.  Lowering the pressure induces further distortions from the ideal octahedral coordination environment in both the H-P-H angles (89.9-90.1$^{\circ}$) and P-H distances (1.437/1.485/1.580~\AA{}).

For the S$_3$PH$_{12}$ (S$_{0.75}$P$_{0.25}$H$_3$) stoichiometry, two metastable cage-like structures with $Imm2$ (100-190~GPa) and $Immm$ (200~GPa) symmetries were found with a distance of 28.4, 18.5, 33.2 meV/atom above the convex hull at 100, 150, and 200~GPa, respectively. The $Immm$ phase was also identified at 200~GPa by Tsuppayakorn-aek et al.\ \cite{Tsuppayakorn:2021}, but the $Imm2$ was not reported before. At 150~GPa the S-H distances ranged from 1.386-1.711~\AA{}, and increasing pressure decreased this range  (1.422-1.613~\AA{} at 200~GPa). The geometry about the P atom deviated somewhat from that of a perfect octahedron with four short and two long P-H distances (1.417 and 1.483~\AA{} at 200~GPa, and  1.422~\AA{}, 1.476 and 1.588~\AA{} at 150~GPa). The DOS at $E_F$ of these two metallic phases was somewhat lower than for 12.5\% doping.

Several other dynamically stable cage-like structures with larger P-dopings were also found: $P1$ S$_3$P$_2$H$_{15}$ (stable between 150-200~GPa), $Amm2$~SPH$_6$ (stable between 100-200~GPa), and $Pm\overline{3}m$ SPH$_2$ (stable at 200~GPa). Their distance from the convex hull was typically somewhat larger than that of the aforementioned phases (Table S2), but still within the realm of synthesizability. Therefore, their structural peculiarities and electronic structures are also briefly described. Whereas, $P1$ S$_3$P$_2$H$_{15}$ can still be viewed as a perturbation of a doped H$_3$S parent, larger phosphorous content has a profound impact on the geometries of the most stable structural alternatives. For a 1:1 chalcogen to pnictogen ratio CSP identified an $Amm2$~SPH$_6$ phase that was built from a 2D square S-H net where the sulfur atoms lay on the corners, and the hydrogen atoms bisected the edges. A layer of PH$_4$ units separated the layers, with the phosphorous atoms forming additional bonds to two sulfur atoms, one in each of the adjacent layers, thereby attaining quasi-octahedral coordination. This phase possessed the lowest enthalpy between 100-200~GPa, and it was 28~meV/atom more stable than the $Imma$ SPH$_6$ structure proposed by Tsuppayakorn-aek and co-workers \cite{Tsuppayakorn:2021}. It is instructive to compare the geometry of $Amm2$~SPH$_6$ with CSH$_7$ \cite{Cui:2020,Sun:2020}, since the two systems have the same valence electron counts. The family of dynamically stable CSH$_7$ structures could also be described as being built from 2D S-H square nets separated by CH$_4$ molecules. But, the smaller radius of carbon relative to phosphorous facilitated the bonding of two adjacent S-H layers via the extra hydrogen atom per formula unit, thereby forming an SH$_3$ cage within which methane was intercalated. 

Finally, a $Pm\overline{3}m$ SPH$_2$ structure whose unit cell contained a phosphorous atom at the center and faces of the cube, and sulfur atoms at the cube edges and vertices was identified at 200~GPa. The phosphorus atom at the center of the cube was coordinated to eight hydrogen atoms at a distance of 1.593~\AA{}. At lower pressures the most stable EA structures consisted of SP layers terminated and separated by atomic or molecular hydrogen, similar to some of the PH$_n$ phases previously proposed between 40-100~GPa \cite{Zurek:2017c}.

$Im\overline{3}m$ H$_3$S is a highly symmetric structure where each H-S distance measures 1.492~\AA{} at 200~GPa because of bond symmetrization that occurs under increasing pressure \cite{Zurek:2016b}. At relatively low levels of $\text{S}\rightarrow \text{P}$ substitutions the doped structures retain the same basic lattice, with slight distortions that can be traced back to the atomic size and electronegativity differences between the two $p$-group elements. With the exception of 12.5\% doping, the DOS at $E_F$ of all of the aforementioned cage-like structures was found to be smaller than that of the H$_3$S parent. Below we will see how chemical substitution affects the propensity for superconductivity of these S-P-H phases.

\subsection{Structures with 2D Hexagonal Sheets}

\begin{figure}
\begin{center}
\includegraphics[width=0.8\columnwidth]{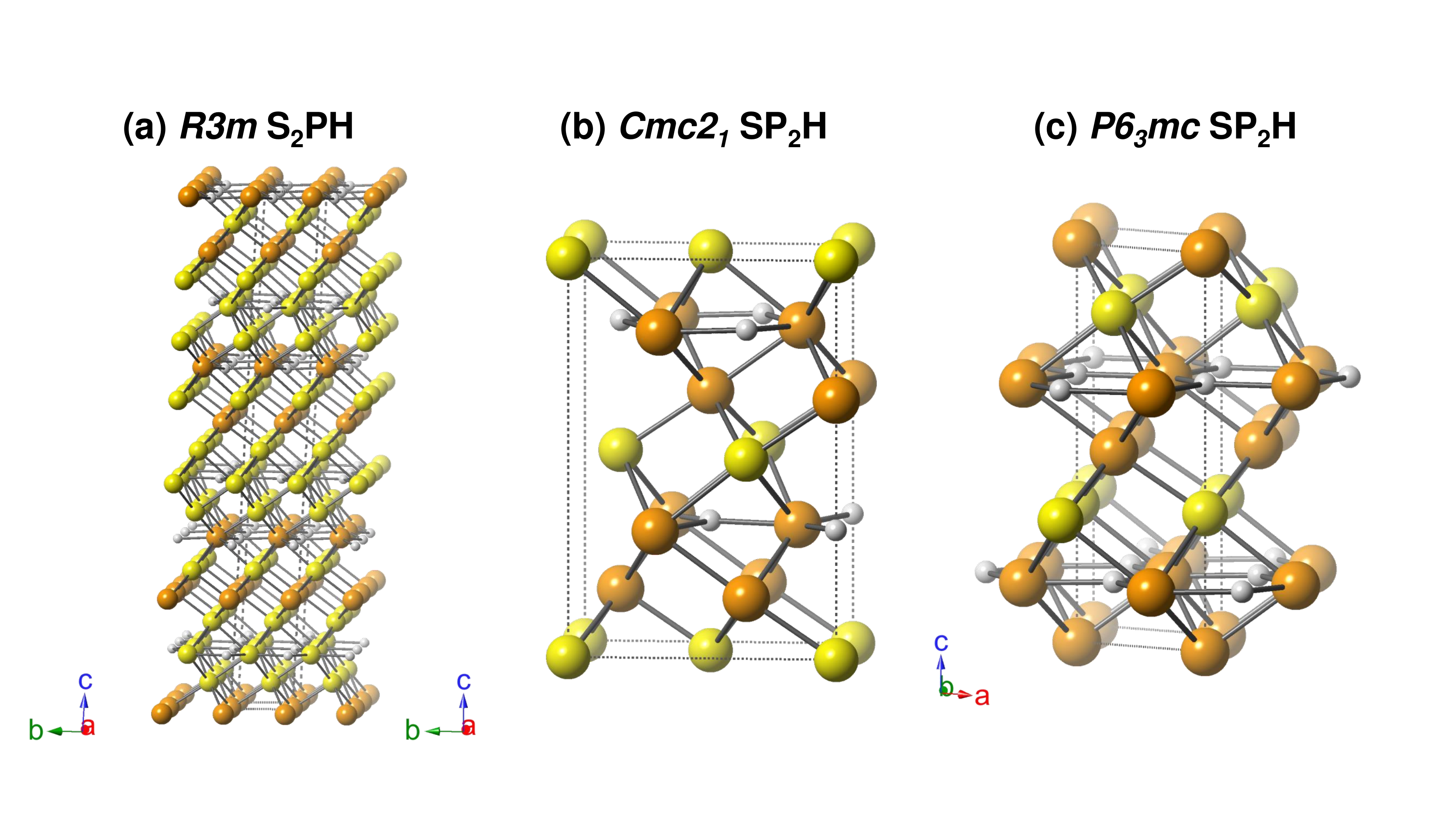}
\end{center}
\caption{Illustrations of predicted S-P-H phases that contained hexagonal H-S and/or P-H sheets: (a) $R3m$ S$_2$PH at 200~GPa, (b) $Cmc2_1$ SP$_2$H at 150~GPa, and (c) $P6_3mc$ SP$_2$H at 200~GPa. S/P/H atoms are colored yellow/orange/white.}  
\label{fig:structure-hexagonal}
\end{figure} 

Stoichiometries that differed from those that could be derived by doping H$_3$S with phosphorous were also explored via evolutionary crystal structure searches. Previous theoretical studies pinpointed a unique $P\bar{6}m2$ CaSH$_3$ phase consisting of alternating CaH$_2$ and honeycomb H-S sheets that possessed two vHs bracketing $E_F$ thereby increasing the number of states that can participate in the pairing mechanism, and concomitantly the $T_c$ of this phase \cite{Yan:2020}.  This structure was metastable down to 128~GPa, where its $T_c$ was estimated to be as high as 100~K. Analogous layered structures were identified in this study (Fig.\ \ref{fig:structure-hexagonal}), but they could be comprised of either H-P or H-S honeycomb sheets, and some contained both.  

For example, $R3m$ S$_2$PH could be described as a layered structure comprised of H-S and H-P hexagonal nets that were separated from each other by a layer of sulfur atoms (located on top of the centers of the hexagons), or by three S-P-S layers. This structure was just 2.7~meV/atom shy of the 200~GPa convex hull.  Whereas the S-H honeycomb nets did not bond with atoms in the adjacent CaH$_2$ layers in $P\overline{6}m2$ CaSH$_3$ \cite{Yan:2020}, here covalent bonds are formed between the main group elements in the honeycomb sheets and the sulfur atoms in the next layer. Both the H-S and H-P bond lengths in the hexagonal sheets present in $R3m$ S$_2$PH (1.650 and 1.658~\AA{}) were longer than the one found in $P\overline{6}m2$ CaSH$_3$  (1.605~\AA{}) at 200~GPa.  At lower pressures the most stable phases with this stoichiometry did not contain these types of hexagonal layer motifs.

Finally, two different SP$_2$H structures, both comprised of 2D hexagonal P-H sheets that were separated by a layer of sulfur and a layer of phosphorous atoms lying above or below the hexagonal holes, were found. The main difference between these two phases is that the $Cmc2_1$ structure, stable between 110-170~GPa, possessed two distinct P-H distances (1.643/1.730~\AA{}), whereas all of the P-H distances were equalized in the higher pressure $P6_3/mc$ phase (1.660~\AA{}). These systems lay 22.4 and 54.8~meV/atom above the convex hull at 150 and 200~GPa, respectively.

\subsection{Structures with 2D Sheets and/or 1D Molecular Chains}

\begin{figure}[!ht]
\begin{center}
\includegraphics[width=0.7\columnwidth]{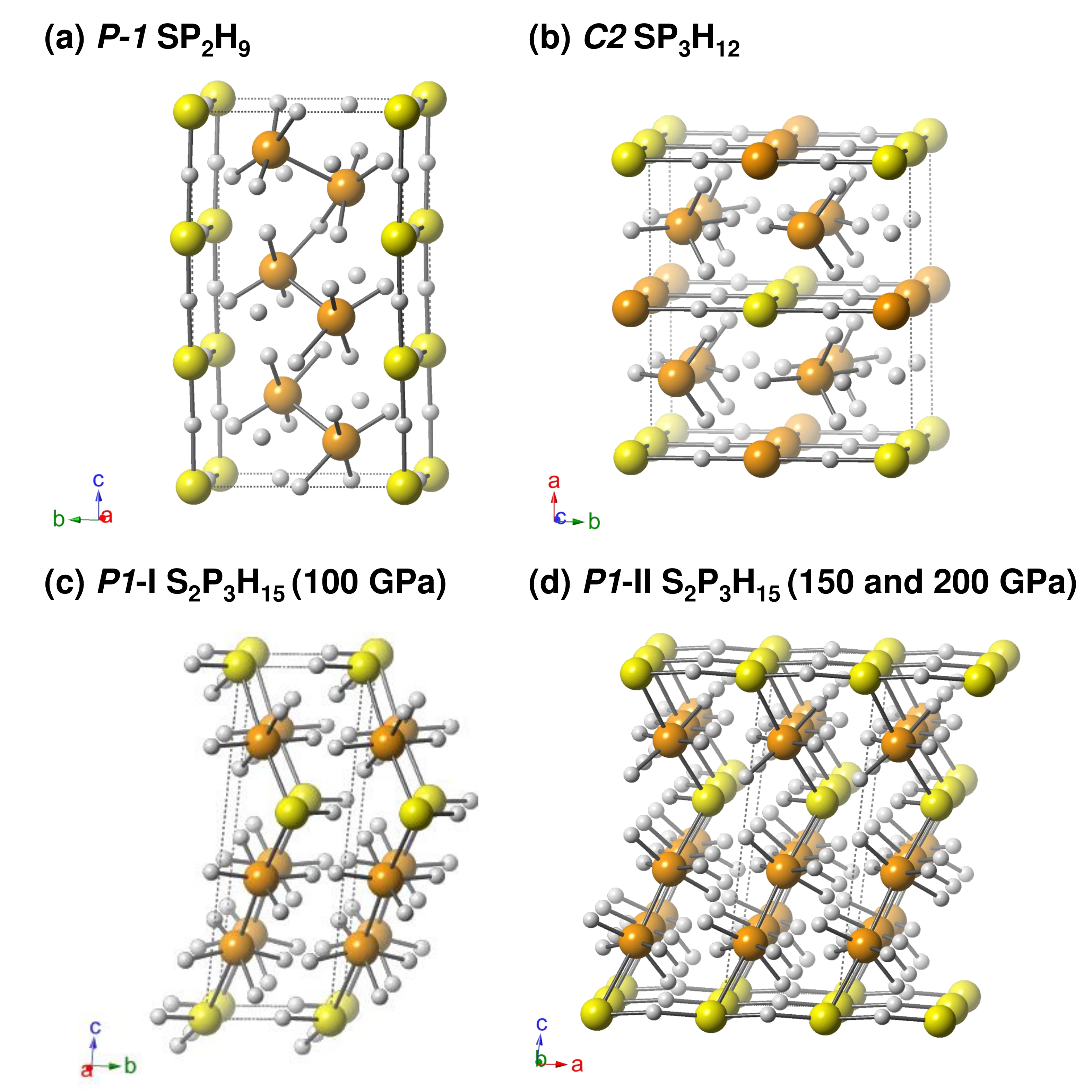}
\end{center}
\caption{Illustrations of predicted S-P-H phases that contained 2D sheets and/or 1D ``nanothreads'': (a) $P\overline{1}$ (S$_{0.33}$P$_{0.67}$H$_3$) SP$_2$H$_9$ at 150~GPa, (b) $C2$ SP$_3$H$_{12}$ at 100~GPa, (c) $P1$-I S$_2$P$_3$H$_{15}$ at 100~GPa, (d) $P1$-II S$_2$P$_3$H$_{15}$ at 150~GPa. S/P/H atoms are colored yellow/orange/white.}  
\label{fig:structure-2Dsheets}
\end{figure} 

In addition to the previously discussed cage-like structures and 2D hexagonal sheets, several metastable phases that contained 2D sheets and/or 1D molecular chains were found in our evolutionary searches. Some examples, illustrated in Fig.\ \ref{fig:structure-2Dsheets}, include SP$_2$H$_{9}$ and S$_2$P$_3$H$_{15}$ (at 150 and 200~GPa), which contained both 1D molecular chains and 2D sheets, SP$_3$H$_{12}$, which possessed 2D sheets and S$_2$P$_3$H$_{15}$ (at 100~GPa), comprised of 1D molecular chains.  Other systems not discussed here (Fig. S42), such as SPH$_5$ and SP$_4$H$_{15}$, only contained 1D molecular chains. 

Though phases with the SP$_2$H$_{9}$ stoichiometry were enthalpically unstable with respect to elemental H$_2$, S and P, their $\Delta H_F$ was still within the threshold proposed to gauge the synthesizability of a metastable material at 1~atm \cite{materialsproject}. At 100 and 150~GPa the most stable SP$_2$H$_{9}$ geometry adopted a $P\bar{1}$ symmetry structure that contained 2D H-S square nets, where the sulfur atoms comprised the edges and the hydrogen atoms the vertices, extending along the $c$-axis. These sheets were separated from one another by a layer of H$_3$P-PH$_3$ molecular motifs that were linked to each other via an extra hydrogen atom bonded to two phosphorous atoms from adjacent units thereby forming a 1D chain analogous to a nanothread \cite{Badding:2017a}. The bond length between phosphorous and a bridging hydrogen atom was longer than between phosphorous and a terminal hydrogen atom (1.51-1.54~\AA{} vs.\ 1.36-1.40~\AA{} at 150~GPa). At 200~GPa the preferred SP$_2$H$_{9}$ structure did not possess any distinguishing features, so it is not discussed here.

Like  SP$_2$H$_{9}$, the most stable SP$_3$H$_{12}$ phases were within the realm of synthesizability even though their enthalpies of formation were positive. At 200~GPa the enthalpy of the most stable structure found at 200~GPa, $Cm$ SP$_3$H$_{12}$, was 33~meV/atom lower than that of a previously proposed phase with this stoichiometry, $Immm$ SP$_3$H$_{12}$ at 200~GPa\cite{Tsuppayakorn:2021}. However, neither it nor the $Cc$ phase found at lower pressures possessed characteristic structural motifs. Therefore, here we describe the $C2$ phase, which was the ground state structure at 100~GPa. $C2$ SP$_2$H$_{9}$ possessed 2D square nets similar to those found within the previously described $P\bar{1}$ SP$_2$H$_{9}$ phase, except some of the sulfur atoms were replaced by phosphorous. These layers were separated by PH$_4$ motifs that were within bonding distance to the $p$-block elements above or below them in the 2D nets. 

 The most stable S$_2$P$_3$H$_{15}$ phase found at 100~GPa can be described as an S-P-H nanothread where each phosphorus atom is octahedrally coordinated to four hydrogen atoms that lie in the same plane and two SH motifs, or to one SH motif and one hydrogen atom.  The sulfur atoms in these SH motifs and the hydrogen atoms are bonded to a nearby PH$_4$ unit. Higher pressures push adjacent nanothreads close enough to form S-H-S bonds along the $a$ axis resulting in a 2D S-H square net that is characteristic of many of these compounds by 150~GPa. Another difference between the preferred S$_2$P$_3$H$_{15}$ structures at 150 and 200~GPa is that the hydrogen atom that bridges two phosphorous atoms in the lower pressure phase migrates to the 2D S-H net upon compression. 

\subsection{Phases on the Hull}

 \begin{figure*}
\begin{center}
\includegraphics[width=0.7\columnwidth]{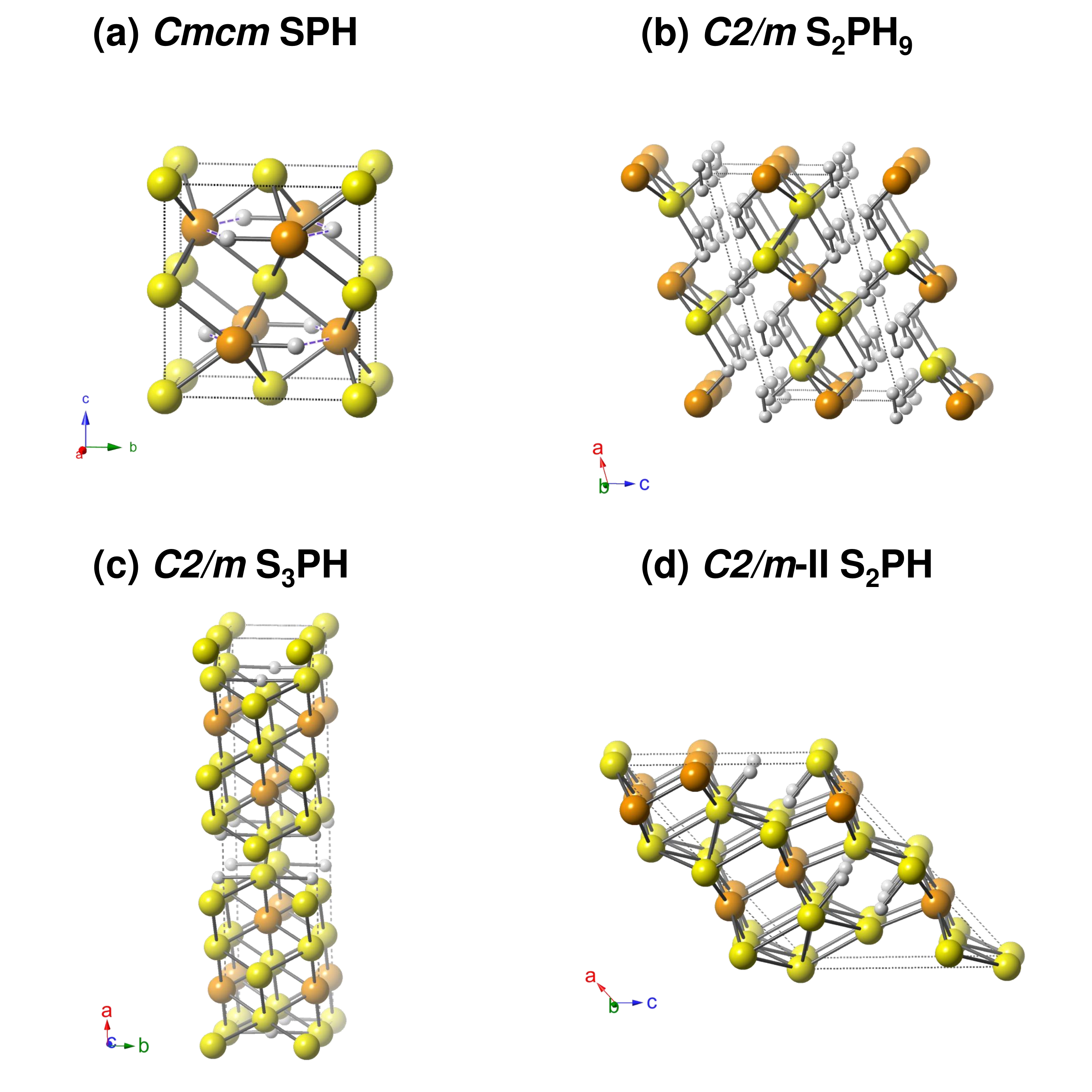}
\end{center}
\caption{Illustrations of predicted S-P-H phases that lay on the convex hull: (a) $Cmcm$ SPH at 200~GPa (Some P-H are connected by dash lines for better eye-view) , (b) $C2/m$ S$_2$PH$_9$ at~100~GPa, (c) $C2/m$ S$_3$PH at 150~GPa, (d) $C2/m$-II S$_2$PH at 150~GPa. S/P/H atoms are colored yellow/orange/white.}  
\label{fig:hullstructures}
\end{figure*}

Our calculations revealed four thermodynamically stable species that lay on the convex hull; these are illustrated in Fig.\ \ref{fig:hullstructures}. A $Cmcm$ symmetry SPH phase that was 5.8~meV/atom above the hull at 150~GPa and lay on the hull at 200~GPa possessed. It can be considered as a distorted ZrBeSi-type lattice (e.g.\ BaAgAs \cite{Mardanya:2019}) with the P and H atoms on the Be and Si sites, but the P and H atoms did not form the hexagonal sheets since the P-H distance (1.459/1.728\AA{}) were too long. At lower pressures a $P2_1/m$ symmetry structure, which is not discussed here because it did not present any distinguishing characteristics, was the most stable SPH stoichiometry phase. 

S$_2$PH$_{9}$ was the only thermodynamically stable system with a S$_x$P$_{1-x}$H$_3$ stoichiometry ($x=0.67$). It lay on the 100~GPa convex hull, but was 18.6/56.9~meV/atom above the hull at 150/200~GPa. At 100~GPa the preferred structure possessed $C2/m$ symmetry, and it could be derived from adding hydrogen to the most stable S$_2$P phase found at this pressure \cite{Liu:2019}. A 1D chain of vertex sharing S-P-S-P rhombii comprised this phase. The phosphorous atoms, which belonged to two rhombii, were octahedrally coordinated to four sulfur atoms and two hydrogen atoms. Each sulfur atom was bonded to a hydrogen atom, which was further bonded to a sulfur atom in an adjacent 1D chain, and H$_2$ molecules were interspersed between the chains.  A $C2/m$ symmetry S$_3$PH phase lay on the 100 and 150~GPa convex hulls, but it was 25.4~meV/atom above the 200~GPa hull. It could be viewed as a distorted cubic network where phosphorous or sulfur atoms were found on the vertices of the cubes. All of the phosphorus atoms were octahedrally coordinated by bonds to sulfur atoms. Many sulfur atoms were octahedrally coordinated as well, and those that weren't were bonded to hydrogen atoms that were shared between two sulfur atoms. Finally, $C2/m$-II S$_2$PH was also thermodynamically stable at 150~GPa.  It resembled S$_3$PH, except the octahedrally coordinated phosphorous  atoms were bonded to both P and S, some of the sides of the distorted cubes were more like parallelograms, and the hydrogen atoms were only bonded to a single sulfur atom. Both S$_2$PH and S$_3$PH could be derived from the most stable high pressure S$_2$P and S$_3$P phases that were  predicted in previous reports \cite{Liu:2019}.

\subsection{Electronic Structure and Superconducting Properties}
The PBE electronic band structures and DOS plots of both thermodynamically and dynamically stable S-P-H phases (Figs.\ S8-26) showed that most were metallic except for $P\bar{1}$ SPH$_4$ at~100 GPa, $Pm$ SPH$_5$ at 100-150~GPa, $P1$ S$_4$PH$_{15}$	(S$_{0.8}$P$_{0.2}$H$_3$) at 100~GPa, $C2/m$ S$_2$PH$_{9}$	(S$_{0.67}$P$_{0.33}$H$_3$) at 100-200~GPa and $P\bar{1}$ S$_2$P$_3$H$_{15}$ (S$_{0.4}$P$_{0.6}$H$_3$) at 100~GPa. In general the DOS at the Fermi level in the SPH$_n$ ($n=1-2$) phases was dominated by the phosphorous and sulfur p states in nearly equivalent proportions. In most of the phases with a larger hydrogen content such as SPH$_n$ ($n=3-6$) all three atom types contributed nearly equally to the DOS at E$_F$, and the most abundant p block element possessed the largest amount of character at $E_F$ in S$_x$P$_y$H ($x,y=2,3$) phases. For the cage-like doped systems illustrated in Fig.\ \ref{fig:structure-cage} the states at $E_F$ were primarily characterized by sulfur and hydrogen character, whereas those with 2D or 1D motifs (i.e.\ those in Fig.\ \ref{fig:structure-2Dsheets}) possessed more hydrogenic character.

A large DOS at $E_F$ is a prerequisite for high $T_c$, in particular if it results from states related to hydrogen. Another descriptor that can be associated with high temperature conventional superconductors is an increased Debye frequency, which is a consequence of the vibrations of atoms with light masses. Therefore, these quantities were calculated for the metallic S-P-H phases found in our work (Table S3-4) and used to screen for the most likely structures to have a high $T_c$. If the DOS per valence electron at $E_F$ was calculated to be $\gtrsim$~0.035 states/eV and the Debye frequency was $\gtrsim$~750~cm$^{-1}$, electron phonon coupling (EPC) calculations were performed to probe the superconducting properties of the phase. In addition, EPC calculations were carried out for SPH, SP$_2$H, S$_2$PH, S$_3$PH, SPH$_2$ and SPH$_3$ though they did not meet these criteria because their superconducting characteristics could be compared to those previously computed for S-P phases. 

Table \ref{tab:Tc-SPH} lists the electron phonon coupling parameter, $\lambda$, and logarithmic average of phonon frequencies ($\omega_\text{log}$) for the considered phases, as well as their $T_c$s as estimated via the Allen-Dynes modified McMillan equation. Comparison of our results with the $T_c$s previously computed \cite{Liu:2019} for SP, SP$_2$, and PS$_2$ (1.5-4.3~K) shows that addition of a single hydrogen atom can raise $T_c$ by an order of magnitude (cf.\ 17-21~K for SPH and SP$_2$H) primarily because of an increase in $\lambda$. Adding one or two more hydrogen atoms, e.g.\ to form SPH$_2$ or SPH$_3$, does not impact $T_c$ by much, but the $T_c$ of $P2_1/m$ SPH$_4$ (Fig.\ S42) was estimated to be significantly larger, 64~K at 200~GPa.

Generally speaking, phases that were constructed by doping $Im\bar{3}m$ H$_3$S with phosphorous, e.g.\ those with S$_{1-x}$P$_{x}$H$_3$ stoichiometries, were the best superconductors.  The $T_c$ of the most stable 12.5\% doped structure at 150~GPa was estimated to be 117~K, which is lower than previous reports for this stoichiometry at the same pressure (168~K \cite{nakanishi:2018} and 190~K \cite{Durajski:2018}). However, the structures considered in prior studies were not found using crystal structure prediction techniques and were at least 7~meV/atom less stable than the phase found here at 150~GPa. The $I4/m$ symmetry 20\% doped structure possessed the second highest $T_c$ of any phase we considered (136~K at 200~GPa), decreasing slightly at lower pressures (114~K at 150~GPa within the $I4$ spacegroup). Because the modified McMillan equation is known to underestimate $T_c$ for systems with large EPC values, we also numerically solved the Eliashberg equations for both the 12.5\% and 20\% dopings. However, comparison of our calculated $T_c$ for $Im\bar{3}m$ H$_3$S at 200~GPa with experiment ($\sim$175-190~K) \cite{Minkov:2020a} shows that when $\mu^*$ is chosen to be 0.1, the modified McMillan equation gives values that better agree with those that were measured.

Increasing the doping to 25\% decreases $T_c$ to 28/93~K at 150/200~GPa. The drop in $T_c$ with an increase in doping correlates with the decreased DOS at $E_F$, which is maximal for 12.5\% doping.  Previously,  Durajski and Szcze\'sniak used the supercell approach to propose a 50\% doped structure with a $T_c$ of 157~K at 155~GPa \cite{Durajski:2018}, while Tsuppayakornaek \emph{et al.} used crystal structure prediction to find an SPH$_6$ phase with an estimated $T_c$ of 89~K at 200~GPa \cite{Tsuppayakorn:2021}. The most stable structure \textsc{XtalOpt} found with this stoichiometry was  70.2 meV/atom lower than the one generated via the supercell approach at 150~GPa, and 28~meV/atom lower than Tsuppayakornaek's structure at 200~GPa. However, the DOS at $E_F$ of our predicted SPH$_{6}$ phase was so low that an EPC calculation was not considered. Even though the stoichiometry of $Pm$ SP$_2$H$_9$ is compatible with that of a P-doped H$_3$S structure, the large concentration of phosphorous has a dramatic impact on the geometry of this phase (Fig. S42), which did not show any distinct structural features. Nonetheless, it's estimated $T_c$, 66~K at 200~GPa, was comparably high.

\begin{table}[ht!]
	\caption{Electron-phonon coupling parameter ($\lambda$), logarithmic average of phonon frequencies ($\omega_\text{log}$), and estimated superconducting critical temperature ($T_c$) for a given pressure of various S-P-H phases. The calculations employed the Allen-Dynes modified McMillan equation with a renormalized Coulomb potential, $\mu^*=0.1$. Values in parentheses were calculated via numerical solution of the Eliashberg equations.} 
	\centering % center the table
	\begin{tabular}{lcccc} % alignment of each column data
    \hline\hline
    System  & Pressure & $\lambda$ & $\omega_\text{log}$ (K) & $T_c$ (K)   \\
    \hline
	$P2_1/m$ SP $^a$	&100		 & 0.46 & 522.4	&  4.3   \\
	$C2/m$ SP$_2$ $^a$	& 150  & 0.38 & 506.8 & 1.5 	\\
    $C2/m$ PS$_2$ $^a$	&	100	 & 0.45 & 453.3 & 1.7  \\
     $P\overline{3}m1$ PS$_2$ $^a$  	&	200	 & 0.37 & 611.0 & 1.5  \\
     \hline
     $R3m$ H$_3$S $^b$ & 130 & 2.07 & 1125.1 & 166 \\
$Im\overline{3}m$ H$_3$S  & 200 & 1.94 & 1337.0 & 188.5 (249.2) \\
\hline
	$Cmcm$ SPH	&	200		& 0.66 & 724.9 & 21.9  \\
		$Cmc2_1$ SP$_2$H	&	150	 & 0.64	& 588.8 & 16.5  \\
		$C2/m$-II S$_2$PH	&	150	&	0.70 & 594.3 & 21.0  \\
  $C2/m$  S$_3$PH & 150  & 0.53 & 562.7&   8.7  \\
    $Cm$  SPH$_2$ & 150  & 0.63 & 739.0 &   19.5  \\
    $P2_1/m$  SPH$_3$ & 150  & 0.57 & 853.3 &   17.0  \\
    $P2_1/m$  SPH$_4$ & 200  & 0.88 & 1124.5 &  63.7  \\
    \hline
      $Cm$ S$_7$PH$_{24}$	(S$_{0.875}$P$_{0.125}$H$_3$) & 100  & 1.06 & 782.4 & 59.4  \\
  $Cm$ S$_7$PH$_{24}$	(S$_{0.875}$P$_{0.125}$H$_3$) & 150  & 1.42 & 1081.2 & 116.8  (147.6)\\
        $Im\overline{3}m$ S$_7$PH$_{24}$	(S$_{0.875}$P$_{0.125}$H$_3$) & 200  & 2.21 & 1188.1 & 182.5  (257.5) \\
  $I4$ S$_4$PH$_{15}$	(S$_{0.8}$P$_{0.2}$H$_3$) & 150  & 1.47 & 1020.5 & 113.9  (145.9)\\
  $I4/m$ S$_4$PH$_{15}$	(S$_{0.8}$P$_{0.2}$H$_3$) & 200 & 1.42 & 1259.2 & 135.9  (171.6) \\
  $Imm2$  S$_3$PH$_{12}$	(S$_{0.75}$P$_{0.25}$H$_3$) & 150  & 0.64 & 1153.0 & 27.3  \\
  $Immm$ S$_3$PH$_{12}$	(S$_{0.75}$P$_{0.25}$H$_3$) & 200  & 1.08 & 1187.5 & 92.6  \\
    $Pm$ SP$_2$H$_{9}$	(S$_{0.33}$P$_{0.67}$H$_3$) & 200  & 0.99 & 951.2 & 65.5  \\
    \hline 
    \end{tabular} \\
    $^a$ Calculated in Ref.\ \cite{Liu:2019}.
    $^b$ Calculated in Ref.\ \cite{Duan:2014}
    \label{tab:Tc-SPH}
\end{table} 

%\subsection{Towards Room Temperature Superconductivity?}

The superconducting critical temperatures of all of the phases discussed so far are lower than that of the parent $Im\bar{3}m$ H$_3$S structure. However, as illustrated in Fig.\ \ref{fig:dos-elf}, at 200~GPa the DOS at $E_F$ of the $Im\bar{3}m$ S$_7$PH$_{24}$ phase is slightly higher than that of the undoped structure. Like sulfur, many textbook examples highlight the propensity of phosphorous to assume hypervalent coordination environments so the phase is stable to lower pressures than octahedrally coordinated carbon in $O_h$-CS$_{15}$H$_{48}$ \cite{Zurek:2021i}. Moreover, unlike in the carbon doped phase the hydrogen atoms retain their multi-centered coordination so the metallicity is not decreased.

Previous studies have computed the $T_c$ of $Im\bar{3}m$ S$_7$PH$_{24}$ and arrived at different conclusions. Durajski and Szcze\'sniak concluded that S$\rightarrow$P replacement cannot raise the $T_c$ of H$_3$S, finding a maximum $T_c$ of 190~K at 155~GPa using the Eliashberg formualism and $\mu^*=0.15$ \cite{Durajski:2018}, but lower level dopings were not considered. 
At about the same time, Nakanishi and co-workers found that 12.5\% doping increases the $T_c$ of H$_3$S at 200~GPa from 189 to 194~K using the Allen-Dynes modified McMillan equation with strong coupling and shape factors, and from 225 to 249~K using the isotropic Eliashberg equation with $\mu^*=0.13$ \cite{nakanishi:2018}. Even higher $T_c$s (212 and 268~K via the Allen-Dynes and Eliashberg methods, respectively) were computed for 6.5\% doping levels. Most recently, Guan and colleagues reported that 12.5\% phosphorous content barely changes the $T_c$ of H$_3$S at 200~GPa \cite{Guan:2021a}. A decreased doping of 6.5\%, on the other hand, increased the $T_c$ of at 200~GPa from 214 to 231~K using the Allen-Dynes modified McMillan equation with strong coupling and shape factors, and from 238 to 262~K via Migdal-Eliashberg theory with $\mu^*=0.1$ \cite{Guan:2021a}. 

Our results echo those of Guan and co-workers: at 200~GPa the estimated $T_c$ of $Im\bar{3}m$ S$_7$PH$_{24}$ differs by 6-8~K from the value predicted for $Im\bar{3}m$ H$_3$S, which is within our calculation error. Therefore, we conclude that the 12.5\% doping of H$_3$S by phosphorous does not change $T_c$ by much. While Guan and Nakanishi both found that 6.25\% doping increased the $T_c$ relative to that of $Im\bar{3}m$ H$_3$S slightly, this system was not considered here because its DOS at $E_F$ was calculated to be lower than that of $Im\bar{3}m$ S$_7$PH$_{24}$ at 200~GPa (Fig. S7).

\section{Conclusions}
The exciting discovery of novel hydride-based superconductors with record breaking superconducting critical temperatures, $T_c$s, has fueled the search for evermore complex systems. One way to design new superconducting materials is by doping known structures in the hopes of finding a material with a higher $T_c$, or a lower stabilization pressure. Previous studies have consider the effect of phosphorous doping on the $T_c$ of $Im\bar{3}m$ H$_3$S, however they were performed either using the virtual crystal approximation or they only considered phases derived from S$\rightarrow$P replacement within the $Im\bar{3}m$ H$_3$S parent.

To overcome these limitations we have employed crystal structure prediction techniques to explore a wide S$_x$P$_y$H$_z$ composition range. The convex hull was mapped out under pressure and found to consist of $Cmcm$ SPH (200~GPa), $C2/m$ S$_2$PH$_9$ (100~GPa), $C2/m$ S$_3$PH (100, 150~GPa) and $C2/m$-II S$_2$PH (150~GPa). Of these phases, the ones whose $T_c$s were calculated ranged from 9-22~K. In addition, a wide variety of potentially synthesizable compounds were predicted containing unique structural motifs such as one-dimensional nanothreads, two-dimensional square nets or honeycomb sheets, as well as cages. From the cages, those with the highest $T_c$s corresponded to low dopings of  $Im\bar{3}m$ H$_3$S with phosphorous. The propensity for phosphorous to participate in hypervalent bonds coupled to its similar-to-sulfur-size were key to retaining a high density of states at the Fermi level in $Im\overline{3}m$ S$_7$PH$_{24}$	(S$_{0.875}$P$_{0.125}$H$_3$), whose $T_c$ was predicted to be 183~K at 200~GPa. We hope this study inspires the eventual synthesis of these compounds, and the exploration of their physical and chemical behavior. \\

\textbf{Acknowledgements:}
We acknowledge the U.S. National Science Foundation (DMR-1827815) for financial support. Calculations were performed at the Center for Computational Research at SUNY Buffalo \cite{ccr}. \\

\textbf{Supporting Information:}
The Supporting Information is available free of charge upon request.

\newpage

\providecommand{\latin}[1]{#1}
\makeatletter
\providecommand{\doi}
  {\begingroup\let\do\@makeother\dospecials
  \catcode`\{=1 \catcode`\}=2 \doi@aux}
\providecommand{\doi@aux}[1]{\endgroup\texttt{#1}}
\makeatother
\providecommand*\mcitethebibliography{\thebibliography}
\csname @ifundefined\endcsname{endmcitethebibliography}
  {\let\endmcitethebibliography\endthebibliography}{}


\begin{mcitethebibliography}{99}
\providecommand*\natexlab[1]{#1}
\providecommand*\mciteSetBstSublistMode[1]{}
\providecommand*\mciteSetBstMaxWidthForm[2]{}
\providecommand*\mciteBstWouldAddEndPuncttrue
  {\def\EndOfBibitem{\unskip.}}
\providecommand*\mciteBstWouldAddEndPunctfalse
  {\let\EndOfBibitem\relax}
\providecommand*\mciteSetBstMidEndSepPunct[3]{}
\providecommand*\mciteSetBstSublistLabelBeginEnd[3]{}
\providecommand*\EndOfBibitem{}
\mciteSetBstSublistMode{f}
\mciteSetBstMaxWidthForm{subitem}{(\alph{mcitesubitemcount})}
\mciteSetBstSublistLabelBeginEnd
  {\mcitemaxwidthsubitemform\space}
  {\relax}
  {\relax}

\bibitem[Ashcroft(1968)]{Ashcroft:1968a}
Ashcroft,~N.~W. Metallic Hydrogen: A High--temperature Superconductor?
  \emph{Phys. Rev. Lett.} \textbf{1968}, \emph{21}, 1748--1749\relax
\mciteBstWouldAddEndPuncttrue
\mciteSetBstMidEndSepPunct{\mcitedefaultmidpunct}
{\mcitedefaultendpunct}{\mcitedefaultseppunct}\relax
\EndOfBibitem
\bibitem[Ashcroft(2004)]{Ashcroft:2004a}
Ashcroft,~N.~W. Hydrogen Dominant Metallic Alloys: High Temperature
  Superconductors? \emph{Phys. Rev. Lett.} \textbf{2004}, \emph{92},
  187002\relax
\mciteBstWouldAddEndPuncttrue
\mciteSetBstMidEndSepPunct{\mcitedefaultmidpunct}
{\mcitedefaultendpunct}{\mcitedefaultseppunct}\relax
\EndOfBibitem
\bibitem[Drozdov \latin{et~al.}(2015)Drozdov, Eremets, Troyan, Ksenofontov, and
  Shylin]{Drozdov:2015}
Drozdov,~A.~P.; Eremets,~M.~I.; Troyan,~I.~A.; Ksenofontov,~V.; Shylin,~S.~I.
  Conventional Superconductivity at 203 Kelvin at High Pressures in the Sulfur
  Hydride System. \emph{Nature} \textbf{2015}, \emph{525}, 73--76\relax
\mciteBstWouldAddEndPuncttrue
\mciteSetBstMidEndSepPunct{\mcitedefaultmidpunct}
{\mcitedefaultendpunct}{\mcitedefaultseppunct}\relax
\EndOfBibitem
\bibitem[Somayazulu \latin{et~al.}(2019)Somayazulu, Ahart, Mishra, Geballe,
  Baldini, Meng, Struzhkin, and Hemley]{Somayazulu:2019}
Somayazulu,~M.; Ahart,~M.; Mishra,~A.~K.; Geballe,~Z.~M.; Baldini,~M.;
  Meng,~Y.; Struzhkin,~V.~V.; Hemley,~R.~J. Evidence for Superconductivity
  above 260~K in Lanthanum Superhydride at Megabar Pressures. \emph{Phys. Rev.
  Lett.} \textbf{2019}, \emph{122}, 027001\relax
\mciteBstWouldAddEndPuncttrue
\mciteSetBstMidEndSepPunct{\mcitedefaultmidpunct}
{\mcitedefaultendpunct}{\mcitedefaultseppunct}\relax
\EndOfBibitem
\bibitem[Drozdov \latin{et~al.}(2019)Drozdov, Kong, Minkov, Besedin,
  Kuzovnikov, Mozaffari, Balicas, Balakirev, Graf, Prakapenka, \latin{et~al.}
  others]{Drozdov:2019}
Drozdov,~A.~P.; Kong,~P.~P.; Minkov,~V.~S.; Besedin,~S.~P.; Kuzovnikov,~M.~A.;
  Mozaffari,~S.; Balicas,~L.; Balakirev,~F.~F.; Graf,~D.~E.; Prakapenka,~V.~B.
  \latin{et~al.}  Superconductivity at 250~K in Lanthanum Hydride Under High
  Pressures. \emph{Nature} \textbf{2019}, \emph{569}, 528--531\relax
\mciteBstWouldAddEndPuncttrue
\mciteSetBstMidEndSepPunct{\mcitedefaultmidpunct}
{\mcitedefaultendpunct}{\mcitedefaultseppunct}\relax
\EndOfBibitem
\bibitem[Troyan \latin{et~al.}(2021)Troyan, Semenok, Kvashnin, Sadakov,
  Sobolevskiy, Pudalov, Ivanova, Prakapenka, Greenberg, Gavriliuk, Lyubutin,
  Struzhkin, Bergara, Errea, Bianco, Calandra, Mauri, Monacelli, Akashi, and
  Oganov]{Troyan-YH4}
Troyan,~I.~A.; Semenok,~D.~V.; Kvashnin,~A.~G.; Sadakov,~A.~V.;
  Sobolevskiy,~O.~A.; Pudalov,~V.~M.; Ivanova,~A.~G.; Prakapenka,~V.~B.;
  Greenberg,~E.; Gavriliuk,~A.~G. \latin{et~al.}  Anomalous High--temperature
  Superconductivity in YH$_6$. \emph{Adv. Mater.} \textbf{2021}, \emph{33},
  2006832\relax
\mciteBstWouldAddEndPuncttrue
\mciteSetBstMidEndSepPunct{\mcitedefaultmidpunct}
{\mcitedefaultendpunct}{\mcitedefaultseppunct}\relax
\EndOfBibitem
\bibitem[Kong \latin{et~al.}(2021)Kong, Minkov, Kuzovnikov, Drozdov, Besedin,
  Mozaffari, Balicas, Balakirev, Prakapenka, Chariton, \latin{et~al.}
  others]{Kong:arxiv}
Kong,~P.; Minkov,~V.~S.; Kuzovnikov,~M.~A.; Drozdov,~A.~P.; Besedin,~S.~P.;
  Mozaffari,~S.; Balicas,~L.; Balakirev,~F.~F.; Prakapenka,~V.~B.; Chariton,~S.
  \latin{et~al.}  Superconductivity up to 243~K in the Yttrium--hydrogen System
  Under High Pressure. \emph{Nat. Commun.} \textbf{2021}, \emph{12}, 5075\relax
\mciteBstWouldAddEndPuncttrue
\mciteSetBstMidEndSepPunct{\mcitedefaultmidpunct}
{\mcitedefaultendpunct}{\mcitedefaultseppunct}\relax
\EndOfBibitem
\bibitem[Snider \latin{et~al.}(2021)Snider, Dasenbrock-Gammon, McBride, Wang,
  Meyers, Lawler, Zurek, Salamat, and Dias]{Snider:2021}
Snider,~E.; Dasenbrock-Gammon,~N.; McBride,~R.; Wang,~X.; Meyers,~N.;
  Lawler,~K.~V.; Zurek,~E.; Salamat,~A.; Dias,~R.~P. Synthesis of Yttrium
  Superhydride Superconductor with a Transition Temperature up to 262~K by
  Catalytic Hydrogenation at High Pressures. \emph{Phys. Rev. Lett.}
  \textbf{2021}, \emph{126}, 117003\relax
\mciteBstWouldAddEndPuncttrue
\mciteSetBstMidEndSepPunct{\mcitedefaultmidpunct}
{\mcitedefaultendpunct}{\mcitedefaultseppunct}\relax
\EndOfBibitem
\bibitem[Snider \latin{et~al.}(2020)Snider, Dasenbrock-Gammon, McBride,
  Debessai, Vindana, Vencatasamy, Lawler, Salamat, and Dias]{Snider:2020}
Snider,~E.; Dasenbrock-Gammon,~N.; McBride,~R.; Debessai,~M.; Vindana,~H.;
  Vencatasamy,~K.; Lawler,~K.~V.; Salamat,~A.; Dias,~R.~P. Room--temperature
  Superconductivity in a Carbonaceous Sulfur Hydride. \emph{Nature}
  \textbf{2020}, \emph{586}, 373--377\relax
\mciteBstWouldAddEndPuncttrue
\mciteSetBstMidEndSepPunct{\mcitedefaultmidpunct}
{\mcitedefaultendpunct}{\mcitedefaultseppunct}\relax
\EndOfBibitem
\bibitem[Zurek and Grochala(2015)Zurek, and Grochala]{Zurek:2014i}
Zurek,~E.; Grochala,~W. Predicting Crystal Structures and Properties of Matter
  Under Extreme Conditions via Quantum Mechanics: The Pressure Is On.
  \emph{Phys. Chem. Chem. Phys.} \textbf{2015}, \emph{17}, 2917--2934\relax
\mciteBstWouldAddEndPuncttrue
\mciteSetBstMidEndSepPunct{\mcitedefaultmidpunct}
{\mcitedefaultendpunct}{\mcitedefaultseppunct}\relax
\EndOfBibitem
\bibitem[Miao \latin{et~al.}(2020)Miao, Sun, Zurek, and Lin]{Zurek:2019k}
Miao,~M.; Sun,~Y.; Zurek,~E.; Lin,~H. Chemistry Under High Pressure. \emph{Nat.
  Rev. Chem.} \textbf{2020}, \emph{508--527}, 4\relax
\mciteBstWouldAddEndPuncttrue
\mciteSetBstMidEndSepPunct{\mcitedefaultmidpunct}
{\mcitedefaultendpunct}{\mcitedefaultseppunct}\relax
\EndOfBibitem
\bibitem[Flores-Livas \latin{et~al.}(2020)Flores-Livas, Boeri, Sanna, Profeta,
  Arita, and Eremets]{Livas:2020-Review}
Flores-Livas,~J.~A.; Boeri,~L.; Sanna,~A.; Profeta,~G.; Arita,~R.; Eremets,~M.
  A Perspective on Conventional High--temperature Superconductors at High
  Pressure: Methods and Materials. \emph{Phys. Rep.} \textbf{2020}, \emph{856},
  1--78\relax
\mciteBstWouldAddEndPuncttrue
\mciteSetBstMidEndSepPunct{\mcitedefaultmidpunct}
{\mcitedefaultendpunct}{\mcitedefaultseppunct}\relax
\EndOfBibitem
\bibitem[Zur()]{Zurek:2018d}
Bi, T.; Zarifi, N.; Terpstra, T.; Zurek, E.~The~Search for Superconductivity in
  High Pressure Hydrides. In $Elsevier$ $Reference$ $Module$ $in$ $Chemistry,$
  $Molecular$ $Sciences$ $and$ $Chemical$ $Engineering$.; Elsevier, 2019; pp
  1--36. DOI:10.1016/B978-0-12-409547-2.11435-0\relax
\mciteBstWouldAddEndPuncttrue
\mciteSetBstMidEndSepPunct{\mcitedefaultmidpunct}
{\mcitedefaultendpunct}{\mcitedefaultseppunct}\relax
\EndOfBibitem
\bibitem[Wang \latin{et~al.}(2018)Wang, Li, Gao, Li, and Ma]{Wang:2018}
Wang,~H.; Li,~X.; Gao,~G.; Li,~Y.; Ma,~Y. Hydrogen-rich Superconductors at High
  Pressures. \emph{Wiley Interdiscip. Rev. Comput. Mol. Sci.} \textbf{2018},
  \emph{8}, e1330\relax
\mciteBstWouldAddEndPuncttrue
\mciteSetBstMidEndSepPunct{\mcitedefaultmidpunct}
{\mcitedefaultendpunct}{\mcitedefaultseppunct}\relax
\EndOfBibitem
\bibitem[Zurek and Bi(2019)Zurek, and Bi]{Zurek:2019h}
Zurek,~E.; Bi,~T. High--temperature Superconductivity in Alkaline and Rare
  Earth Polyhydrides at High Pressure: A Theoretical Perspective. \emph{J.
  Chem. Phys.} \textbf{2019}, \emph{150}, 050901\relax
\mciteBstWouldAddEndPuncttrue
\mciteSetBstMidEndSepPunct{\mcitedefaultmidpunct}
{\mcitedefaultendpunct}{\mcitedefaultseppunct}\relax
\EndOfBibitem
\bibitem[Boeri \latin{et~al.}(2021)Boeri, Hennig, Hirschfeld, Profeta, Sanna,
  Zurek, Pickett, Amsler, Dias, Eremets, \latin{et~al.} others]{Boeri:2021}
Boeri,~L.; Hennig,~R.~G.; Hirschfeld,~P.~J.; Profeta,~G.; Sanna,~A.; Zurek,~E.;
  Pickett,~W.~E.; Amsler,~M.; Dias,~R.; Eremets,~M. \latin{et~al.}  The 2021
  Room--temperature Superconductivity Roadmap. \emph{J. Phys.: Condens. Matter}
  \textbf{2021}, ASAP\relax
\mciteBstWouldAddEndPuncttrue
\mciteSetBstMidEndSepPunct{\mcitedefaultmidpunct}
{\mcitedefaultendpunct}{\mcitedefaultseppunct}\relax
\EndOfBibitem
\bibitem[Hilleke and Zurek(2021)Hilleke, and Zurek]{Zurek:2021k}
Hilleke,~K.~P.; Zurek,~E. Tuning Chemical Precompression: Theoretical Design
  and Crystal Chemistry of Novel Hydrides in the Quest for Warm and Light
  Superconductivity at Ambient Pressures. \emph{arXiv preprint} \textbf{2021},
  arXiv:2111.03697, submitted\relax
\mciteBstWouldAddEndPuncttrue
\mciteSetBstMidEndSepPunct{\mcitedefaultmidpunct}
{\mcitedefaultendpunct}{\mcitedefaultseppunct}\relax
\EndOfBibitem
\bibitem[Wang \latin{et~al.}(2012)Wang, Tse, Tanaka, Iitaka, and Ma]{Wang:2012}
Wang,~H.; Tse,~J.~S.; Tanaka,~K.; Iitaka,~T.; Ma,~Y. Superconductive
  Sodalite--like Clathrate Calcium Hydride at High Pressures. \emph{Proc. Natl.
  Acad. Sci. U.S.A.} \textbf{2012}, \emph{109}, 6463--6466\relax
\mciteBstWouldAddEndPuncttrue
\mciteSetBstMidEndSepPunct{\mcitedefaultmidpunct}
{\mcitedefaultendpunct}{\mcitedefaultseppunct}\relax
\EndOfBibitem
\bibitem[Duan \latin{et~al.}(2014)Duan, Liu, Tian, Li, Huang, Zhao, Yu, Liu,
  Tian, and Cui]{Duan:2014}
Duan,~D.; Liu,~Y.; Tian,~F.; Li,~D.; Huang,~X.; Zhao,~Z.; Yu,~H.; Liu,~B.;
  Tian,~W.; Cui,~T. Pressure--induced Metallization of Dense (H$_2$S)$_2$H$_2$
  with High--$T_c$ Superconductivity. \emph{Sci. Rep.} \textbf{2014}, \emph{4},
  6968\relax
\mciteBstWouldAddEndPuncttrue
\mciteSetBstMidEndSepPunct{\mcitedefaultmidpunct}
{\mcitedefaultendpunct}{\mcitedefaultseppunct}\relax
\EndOfBibitem
\bibitem[Yan \latin{et~al.}(2020)Yan, Bi, Geng, Wang, and Zurek]{Yan:2020}
Yan,~Y.; Bi,~T.; Geng,~N.; Wang,~X.; Zurek,~E. A Metastable CaSH$_3$ Phase
  Composed of HS Honeycomb Sheets That Is Superconducting Under Pressure.
  \emph{J. Phys. Chem. Lett.} \textbf{2020}, \emph{11}, 9629--9636\relax
\mciteBstWouldAddEndPuncttrue
\mciteSetBstMidEndSepPunct{\mcitedefaultmidpunct}
{\mcitedefaultendpunct}{\mcitedefaultseppunct}\relax
\EndOfBibitem
\bibitem[Peng \latin{et~al.}(2017)Peng, Sun, Pickard, Needs, Wu, and
  Ma]{Peng:2017h}
Peng,~F.; Sun,~Y.; Pickard,~C.~J.; Needs,~R.~J.; Wu,~Q.; Ma,~Y. Hydrogen
  Clathrate Structures in Rare Earth Hydrides at High Pressures: Possible Route
  to Room--temperature Superconductivity. \emph{Phys. Rev. Lett.}
  \textbf{2017}, \emph{119}, 107001\relax
\mciteBstWouldAddEndPuncttrue
\mciteSetBstMidEndSepPunct{\mcitedefaultmidpunct}
{\mcitedefaultendpunct}{\mcitedefaultseppunct}\relax
\EndOfBibitem
\bibitem[Liu \latin{et~al.}(2017)Liu, Naumov, Hoffmann, Ashcroft, and
  Hemley]{Liu:2017}
Liu,~H.; Naumov,~I.~I.; Hoffmann,~R.; Ashcroft,~N.~W.; Hemley,~R.~J. Potential
  High--$T_c$ Superconducting Lanthanum and Yttrium Hydrides at High Pressure.
  \emph{Proc. Natl. Acad. Sci. U.S.A.} \textbf{2017}, \emph{114},
  6990--6995\relax
\mciteBstWouldAddEndPuncttrue
\mciteSetBstMidEndSepPunct{\mcitedefaultmidpunct}
{\mcitedefaultendpunct}{\mcitedefaultseppunct}\relax
\EndOfBibitem
\bibitem[Semenok \latin{et~al.}(2021)Semenok, Troyan, Ivanova, Kvashnin,
  Kruglov, Hanfland, Sadakov, Sobolevskiy, Pervakov, Lyubutin, Glazyrin,
  Giordano, Karimov, Vasiliev, Akashi, Pudalov, and Oganov]{Semenok:2021b}
Semenok,~D.~V.; Troyan,~I.~A.; Ivanova,~A.~G.; Kvashnin,~A.~G.; Kruglov,~I.~A.;
  Hanfland,~M.; Sadakov,~A.~V.; Sobolevskiy,~O.~A.; Pervakov,~K.~S.;
  Lyubutin,~I.~S. \latin{et~al.}  Superconductivity at 253~K in
  Lanthanum--yttrium Ternary Hydrides. \emph{Mater. Today} \textbf{2021},
  \emph{48}, 18--28\relax
\mciteBstWouldAddEndPuncttrue
\mciteSetBstMidEndSepPunct{\mcitedefaultmidpunct}
{\mcitedefaultendpunct}{\mcitedefaultseppunct}\relax
\EndOfBibitem
\bibitem[Sun \latin{et~al.}(2019)Sun, Lv, Xie, Liu, and Ma]{Sun:2019}
Sun,~Y.; Lv,~J.; Xie,~Y.; Liu,~H.; Ma,~Y. Route to a Superconducting Phase
  above Room Temperature in Electron--doped Hydride Compounds Under High
  Pressure. \emph{Phys. Rev. Lett.} \textbf{2019}, \emph{123}, 097001\relax
\mciteBstWouldAddEndPuncttrue
\mciteSetBstMidEndSepPunct{\mcitedefaultmidpunct}
{\mcitedefaultendpunct}{\mcitedefaultseppunct}\relax
\EndOfBibitem
\bibitem[Lonie \latin{et~al.}(2013)Lonie, Hooper, Altintas, and
  Zurek]{Zurek:2012m}
Lonie,~D.~C.; Hooper,~J.; Altintas,~B.; Zurek,~E. Metallization of Magnesium
  Polyhydrides Under Pressure. \emph{Phys. Rev. B} \textbf{2013}, \emph{87},
  054107\relax
\mciteBstWouldAddEndPuncttrue
\mciteSetBstMidEndSepPunct{\mcitedefaultmidpunct}
{\mcitedefaultendpunct}{\mcitedefaultseppunct}\relax
\EndOfBibitem
\bibitem[Di~Cataldo \latin{et~al.}(2021)Di~Cataldo, Heil, von~der Linden, and
  Boeri]{diCataldo:2021}
Di~Cataldo,~S.; Heil,~C.; von~der Linden,~W.; Boeri,~L. LaBH$_8$: Towards
  High--$T_c$ Low--pressure Superconductivity in Ternary Superhydrides.
  \emph{Phys. Rev. B} \textbf{2021}, \emph{104}, L020511\relax
\mciteBstWouldAddEndPuncttrue
\mciteSetBstMidEndSepPunct{\mcitedefaultmidpunct}
{\mcitedefaultendpunct}{\mcitedefaultseppunct}\relax
\EndOfBibitem
\bibitem[Liang \latin{et~al.}(2021)Liang, Bergara, Wei, Wang, Sun, Liu, Hemley,
  Wang, Gao, and Tian]{Liang:2021}
Liang,~X.; Bergara,~A.; Wei,~X.; Wang,~L.; Sun,~R.; Liu,~H.; Hemley,~R.~J.;
  Wang,~L.; Gao,~G.; Tian,~Y. Prediction of High--T$_c$ Superconductivity in
  Ternary Lanthanum Borohydrides. \emph{Phys. Rev. B} \textbf{2021},
  \emph{104}, 134501\relax
\mciteBstWouldAddEndPuncttrue
\mciteSetBstMidEndSepPunct{\mcitedefaultmidpunct}
{\mcitedefaultendpunct}{\mcitedefaultseppunct}\relax
\EndOfBibitem
\bibitem[Zhang \latin{et~al.}(2021)Zhang, Cui, Hutcheon, Shipley, Song, Du,
  Kresin, Duan, Pickard, and Yao]{Zhang:2021}
Zhang,~Z.; Cui,~T.; Hutcheon,~M.~J.; Shipley,~A.~M.; Song,~H.; Du,~M.;
  Kresin,~V.; Duan,~D.; Pickard,~C.~J.; Yao,~Y. Design Principles for High
  Temperature Superconductors with Hydrogen--based Alloy Backbone at Moderate
  Pressure. \emph{arXiv preprint} \textbf{2021}, arXiv:2106.09879\relax
\mciteBstWouldAddEndPuncttrue
\mciteSetBstMidEndSepPunct{\mcitedefaultmidpunct}
{\mcitedefaultendpunct}{\mcitedefaultseppunct}\relax
\EndOfBibitem
\bibitem[Heil and Boeri(2015)Heil, and Boeri]{Heil:2015a}
Heil,~C.; Boeri,~L. Influence of Bonding on Superconductivity in High--pressure
  Hydrides. \emph{Phys. Rev. B} \textbf{2015}, \emph{92}, 060508(R)\relax
\mciteBstWouldAddEndPuncttrue
\mciteSetBstMidEndSepPunct{\mcitedefaultmidpunct}
{\mcitedefaultendpunct}{\mcitedefaultseppunct}\relax
\EndOfBibitem
\bibitem[Ge \latin{et~al.}(2016)Ge, Zhang, and Yao]{Ge:2016}
Ge,~Y.; Zhang,~F.; Yao,~Y. First--principles Demonstration of Superconductivity
  at 280~K in Hydrogen Sulfide with Low Phosphorus Substitution. \emph{Phys.
  Rev. B} \textbf{2016}, \emph{93}, 224513\relax
\mciteBstWouldAddEndPuncttrue
\mciteSetBstMidEndSepPunct{\mcitedefaultmidpunct}
{\mcitedefaultendpunct}{\mcitedefaultseppunct}\relax
\EndOfBibitem
\bibitem[Ge \latin{et~al.}(2020)Ge, Zhang, Dias, Hemley, and Yao]{Ge:2020a}
Ge,~Y.; Zhang,~F.; Dias,~R.~P.; Hemley,~R.~J.; Yao,~Y. Hole--doped
  Room--temperature Superconductivity in H$_3$S$_{1-x}$Z$_x$ (Z=C, Si).
  \emph{Mater. Today Phys.} \textbf{2020}, \emph{15}, 100330\relax
\mciteBstWouldAddEndPuncttrue
\mciteSetBstMidEndSepPunct{\mcitedefaultmidpunct}
{\mcitedefaultendpunct}{\mcitedefaultseppunct}\relax
\EndOfBibitem
\bibitem[Hu \latin{et~al.}(2020)Hu, Paul, Karasiev, and Dias]{Hu:2020}
Hu,~S.~X.; Paul,~R.; Karasiev,~V.~V.; Dias,~R.~P. Carbon-doped Sulfur Hydrides
  as Room--temperature Superconductors at 270~GPa. \emph{arXiv preprint}
  \textbf{2020}, arXiv:2012.10259\relax
\mciteBstWouldAddEndPuncttrue
\mciteSetBstMidEndSepPunct{\mcitedefaultmidpunct}
{\mcitedefaultendpunct}{\mcitedefaultseppunct}\relax
\EndOfBibitem
\bibitem[Fan \latin{et~al.}(2016)Fan, Papaconstantopoulos, Mehl, and
  Klein]{Fan:2016}
Fan,~F.; Papaconstantopoulos,~D.~A.; Mehl,~M.~J.; Klein,~B.~M.
  High--temperature Superconductivity at High pressures for
  H$_3$Si$_x$P$_{1-x}$, H$_3$P$_x$S$_{1-x}$, and H$_3$Cl$_x$S$_{1-x}$. \emph{J.
  Phys. Chem. Solids} \textbf{2016}, \emph{99}, 105--110\relax
\mciteBstWouldAddEndPuncttrue
\mciteSetBstMidEndSepPunct{\mcitedefaultmidpunct}
{\mcitedefaultendpunct}{\mcitedefaultseppunct}\relax
\EndOfBibitem
\bibitem[Wang \latin{et~al.}(2021)Wang, Hirayama, Nomoto, Koretsune, Arita, and
  Flores-Livas]{Wang:2021a}
Wang,~T.; Hirayama,~M.; Nomoto,~T.; Koretsune,~T.; Arita,~R.;
  Flores-Livas,~J.~A. Absence of Conventional Room--temperature
  Superconductivity at High Pressure in Carbon--doped H$_3$S. \emph{Phys. Rev.
  B} \textbf{2021}, \emph{104}, 064510\relax
\mciteBstWouldAddEndPuncttrue
\mciteSetBstMidEndSepPunct{\mcitedefaultmidpunct}
{\mcitedefaultendpunct}{\mcitedefaultseppunct}\relax
\EndOfBibitem
\bibitem[Wang \latin{et~al.}(2021)Wang, Bi, Hilleke, Lamichhane, Hemley, and
  Zurek]{Zurek:2021i}
Wang,~X.; Bi,~T.; Hilleke,~K.~P.; Lamichhane,~A.; Hemley,~R.~J.; Zurek,~E. A
  Little Bit of Carbon Can Do a Lot for Superconductivity in H$_3$S.
  \emph{arXiv preprint} \textbf{2021}, arXiv:2109.09898v2\relax
\mciteBstWouldAddEndPuncttrue
\mciteSetBstMidEndSepPunct{\mcitedefaultmidpunct}
{\mcitedefaultendpunct}{\mcitedefaultseppunct}\relax
\EndOfBibitem
\bibitem[Sun \latin{et~al.}(2020)Sun, Tian, Jiang, Li, Li, Iitaka, Zhong, and
  Xie]{Sun:2020}
Sun,~Y.; Tian,~Y.; Jiang,~B.; Li,~X.; Li,~H.; Iitaka,~T.; Zhong,~X.; Xie,~Y.
  Computational Discovery of a Dynamically Stable Cubic SH$_3$--like
  High--temperature Superconductor at 100~GPa via CH$_4$ Intercalation.
  \emph{Phys. Rev. B} \textbf{2020}, \emph{101}, 174102\relax
\mciteBstWouldAddEndPuncttrue
\mciteSetBstMidEndSepPunct{\mcitedefaultmidpunct}
{\mcitedefaultendpunct}{\mcitedefaultseppunct}\relax
\EndOfBibitem
\bibitem[Cui \latin{et~al.}(2020)Cui, Bi, Shi, Li, Liu, Zurek, and
  Hemley]{Cui:2020}
Cui,~W.; Bi,~T.; Shi,~J.; Li,~Y.; Liu,~H.; Zurek,~E.; Hemley,~R.~J. Route to
  high--$T_c$ Superconductivity via CH$_4$--intercalated H$_3$S Hydride
  Perovskites. \emph{Phys. Rev. B} \textbf{2020}, \emph{101}, 134504\relax
\mciteBstWouldAddEndPuncttrue
\mciteSetBstMidEndSepPunct{\mcitedefaultmidpunct}
{\mcitedefaultendpunct}{\mcitedefaultseppunct}\relax
\EndOfBibitem
\bibitem[Amsler(2019)]{Amsler:2019}
Amsler,~M. Thermodynamics and Superconductivity of S$_x$Se$_{1-x}$H$_3$.
  \emph{Phys. Rev. B} \textbf{2019}, \emph{99}, 060102(R)\relax
\mciteBstWouldAddEndPuncttrue
\mciteSetBstMidEndSepPunct{\mcitedefaultmidpunct}
{\mcitedefaultendpunct}{\mcitedefaultseppunct}\relax
\EndOfBibitem
\bibitem[Liu \latin{et~al.}(2018)Liu, Cui, Shi, Zhu, Chen, Lin, Su, Ma, Yang,
  Xu, Hao, Durajski, Qi, Li, and Li]{Liu:2018prb}
Liu,~B.; Cui,~W.; Shi,~J.; Zhu,~L.; Chen,~J.; Lin,~S.; Su,~R.; Ma,~J.;
  Yang,~K.; Xu,~M. \latin{et~al.}  Effect of Covalent Bonding on the
  Superconducting Critical Temperature of the H--S--Se System. \emph{Phys. Rev.
  B} \textbf{2018}, \emph{98}, 174101\relax
\mciteBstWouldAddEndPuncttrue
\mciteSetBstMidEndSepPunct{\mcitedefaultmidpunct}
{\mcitedefaultendpunct}{\mcitedefaultseppunct}\relax
\EndOfBibitem
\bibitem[Nakanishi \latin{et~al.}(2018)Nakanishi, Ishikawa, and
  Shimizu]{nakanishi:2018}
Nakanishi,~A.; Ishikawa,~T.; Shimizu,~K. First--principles Study on
  Superconductivity of P--and Cl--doped H$_3$S. \emph{J. Phys. Soc. Jpn.}
  \textbf{2018}, \emph{87}, 124711\relax
\mciteBstWouldAddEndPuncttrue
\mciteSetBstMidEndSepPunct{\mcitedefaultmidpunct}
{\mcitedefaultendpunct}{\mcitedefaultseppunct}\relax
\EndOfBibitem
\bibitem[Guan \latin{et~al.}(2021)Guan, Sun, and Liu]{Guan:2021a}
Guan,~H.; Sun,~Y.; Liu,~H. Superconductivity of H$_3$S Doped with Light
  Elements. \emph{Phys. Rev. Research} \textbf{2021}, \emph{3}, 043102\relax
\mciteBstWouldAddEndPuncttrue
\mciteSetBstMidEndSepPunct{\mcitedefaultmidpunct}
{\mcitedefaultendpunct}{\mcitedefaultseppunct}\relax
\EndOfBibitem
\bibitem[Durajski and Szcz{\c e}{\'s}niak(2018)Durajski, and Szcz{\c
  e}{\'s}niak]{Durajski:2018}
Durajski,~A.~P.; Szcz{\c e}{\'s}niak,~R. Gradual Reduction of the
  Superconducting Transition Temperature of H$_3$S by Partial Replacing Sulfur
  with Phosphorus. \emph{Physica C} \textbf{2018}, \emph{554}, 38--43\relax
\mciteBstWouldAddEndPuncttrue
\mciteSetBstMidEndSepPunct{\mcitedefaultmidpunct}
{\mcitedefaultendpunct}{\mcitedefaultseppunct}\relax
\EndOfBibitem
\bibitem[Tsuppayakorn-aek \latin{et~al.}(2021)Tsuppayakorn-aek, Phansuke,
  Kaewtubtim, Ahuja, and Bovornratanaraks]{Tsuppayakorn:2021}
Tsuppayakorn-aek,~P.; Phansuke,~P.; Kaewtubtim,~P.; Ahuja,~R.;
  Bovornratanaraks,~T. Enthalpy Stabilization of Superconductivity in an
  Alloying S--P--H System: First--principles Cluster Expansion Study Under High
  Pressure. \emph{Comput. Mater. Sci.} \textbf{2021}, \emph{190}, 110282\relax
\mciteBstWouldAddEndPuncttrue
\mciteSetBstMidEndSepPunct{\mcitedefaultmidpunct}
{\mcitedefaultendpunct}{\mcitedefaultseppunct}\relax
\EndOfBibitem
\bibitem[Lonie and Zurek(2011)Lonie, and Zurek]{Zurek:2011a}
Lonie,~D.~C.; Zurek,~E. XtalOpt: An open--source Evolutionary Algorithm for
  Crystal Structure Prediction. \emph{Comput. Phys. Commun.} \textbf{2011},
  \emph{182}, 372--387\relax
\mciteBstWouldAddEndPuncttrue
\mciteSetBstMidEndSepPunct{\mcitedefaultmidpunct}
{\mcitedefaultendpunct}{\mcitedefaultseppunct}\relax
\EndOfBibitem
\bibitem[xta()]{xtalopt-web}
http://xtalopt.github.io/\relax
\mciteBstWouldAddEndPuncttrue
\mciteSetBstMidEndSepPunct{\mcitedefaultmidpunct}
{\mcitedefaultendpunct}{\mcitedefaultseppunct}\relax
\EndOfBibitem
\bibitem[Falls \latin{et~al.}(2021)Falls, Avery, Wang, Hilleke, and
  Zurek]{Zurek:2020i}
Falls,~Z.; Avery,~P.; Wang,~X.; Hilleke,~K.~P.; Zurek,~E. The XtalOpt
  Evolutionary Algorithm for Crystal Structure Prediction. \emph{J. Phys. Chem.
  C} \textbf{2021}, \emph{125}, 1601--1620\relax
\mciteBstWouldAddEndPuncttrue
\mciteSetBstMidEndSepPunct{\mcitedefaultmidpunct}
{\mcitedefaultendpunct}{\mcitedefaultseppunct}\relax
\EndOfBibitem
\bibitem[Avery \latin{et~al.}(2019)Avery, Toher, Curtarolo, and
  Zurek]{Zurek:2018j}
Avery,~P.; Toher,~C.; Curtarolo,~S.; Zurek,~E. XtalOpt Version r12: An
  Open--source Evolutionary Algorithm for Crystal Structure Prediction.
  \emph{Comput. Phys. Commun.} \textbf{2019}, \emph{237}, 274--275\relax
\mciteBstWouldAddEndPuncttrue
\mciteSetBstMidEndSepPunct{\mcitedefaultmidpunct}
{\mcitedefaultendpunct}{\mcitedefaultseppunct}\relax
\EndOfBibitem
\bibitem[Avery and Zurek(2017)Avery, and Zurek]{Zurek:2016h}
Avery,~P.; Zurek,~E. RandSpg: An Open--source Program for Generating Atomistic
  Crystal Structures with Specific Spacegroups. \emph{Comput. Phys. Commun.}
  \textbf{2017}, \emph{213}, 208--216\relax
\mciteBstWouldAddEndPuncttrue
\mciteSetBstMidEndSepPunct{\mcitedefaultmidpunct}
{\mcitedefaultendpunct}{\mcitedefaultseppunct}\relax
\EndOfBibitem
\bibitem[Lonie and Zurek(2012)Lonie, and Zurek]{Zurek:2011i}
Lonie,~D.~C.; Zurek,~E. Identifying Duplicate Crystal Structures: XtalComp, An
  Open--source solution. \emph{Comput. Phys. Commun.} \textbf{2012},
  \emph{183}, 690--697\relax
\mciteBstWouldAddEndPuncttrue
\mciteSetBstMidEndSepPunct{\mcitedefaultmidpunct}
{\mcitedefaultendpunct}{\mcitedefaultseppunct}\relax
\EndOfBibitem
\bibitem[Kresse and Hafner(1993)Kresse, and Hafner]{Kresse:1993a}
Kresse,~G.; Hafner,~J. \textit{Ab initio} Molecular Dynamics for Liquid Metals.
  \emph{Phys. Rev. B.} \textbf{1993}, \emph{47}, 558--561\relax
\mciteBstWouldAddEndPuncttrue
\mciteSetBstMidEndSepPunct{\mcitedefaultmidpunct}
{\mcitedefaultendpunct}{\mcitedefaultseppunct}\relax
\EndOfBibitem
\bibitem[Kresse and Joubert(1999)Kresse, and Joubert]{Kresse:1999a}
Kresse,~G.; Joubert,~D. From Ultrasoft Pseudopotentials to the Projector
  Augmented--wave Method. \emph{Phys. Rev. B.} \textbf{1999}, \emph{59},
  1758--1775\relax
\mciteBstWouldAddEndPuncttrue
\mciteSetBstMidEndSepPunct{\mcitedefaultmidpunct}
{\mcitedefaultendpunct}{\mcitedefaultseppunct}\relax
\EndOfBibitem
\bibitem[Perdew \latin{et~al.}(1996)Perdew, Burke, and Ernzerhof]{Perdew:1996a}
Perdew,~J.~P.; Burke,~K.; Ernzerhof,~M. Generalized Gradient Approximation Made
  Simple. \emph{Phys. Rev. Lett.} \textbf{1996}, \emph{77}, 3865--3868\relax
\mciteBstWouldAddEndPuncttrue
\mciteSetBstMidEndSepPunct{\mcitedefaultmidpunct}
{\mcitedefaultendpunct}{\mcitedefaultseppunct}\relax
\EndOfBibitem
\bibitem[Maintz \latin{et~al.}(2016)Maintz, Deringer, Tchougr{\'e}eff, and
  Dronskowski]{Maintz:2016}
Maintz,~S.; Deringer,~V.~L.; Tchougr{\'e}eff,~A.~L.; Dronskowski,~R. LOBSTER: A
  Tool to Extract Chemical Bonding from Plane--wave Based DFT. \emph{J. Comput.
  Chem.} \textbf{2016}, \emph{37}, 1030--1035\relax
\mciteBstWouldAddEndPuncttrue
\mciteSetBstMidEndSepPunct{\mcitedefaultmidpunct}
{\mcitedefaultendpunct}{\mcitedefaultseppunct}\relax
\EndOfBibitem
\bibitem[Bl{\"o}chl(1994)]{Blochl:1994a}
Bl{\"o}chl,~P.~E. Projector Augmented--wave Method. \emph{Phys. Rev. B}
  \textbf{1994}, \emph{50}, 17953--17979\relax
\mciteBstWouldAddEndPuncttrue
\mciteSetBstMidEndSepPunct{\mcitedefaultmidpunct}
{\mcitedefaultendpunct}{\mcitedefaultseppunct}\relax
\EndOfBibitem
\bibitem[Parlinski \latin{et~al.}(1997)Parlinski, Li, and
  Kawazoe]{Parlinski:1997}
Parlinski,~K.; Li,~Z.~Q.; Kawazoe,~Y. First--principles Determination of the
  Soft Mode in Cubic ZrO$_2$. \emph{Phys. Rev. Lett.} \textbf{1997},
  4063--4066\relax
\mciteBstWouldAddEndPuncttrue
\mciteSetBstMidEndSepPunct{\mcitedefaultmidpunct}
{\mcitedefaultendpunct}{\mcitedefaultseppunct}\relax
\EndOfBibitem
\bibitem[Chaput \latin{et~al.}(2011)Chaput, Togo, Tanaka, and Hug]{Chaput:2011}
Chaput,~L.; Togo,~A.; Tanaka,~I.; Hug,~G. Phonon--phonon Interactions in
  Transition Metals. \emph{Phys. Rev. B} \textbf{2011}, \emph{84}, 094302\relax
\mciteBstWouldAddEndPuncttrue
\mciteSetBstMidEndSepPunct{\mcitedefaultmidpunct}
{\mcitedefaultendpunct}{\mcitedefaultseppunct}\relax
\EndOfBibitem
\bibitem[Togo and Tanaka(2015)Togo, and Tanaka]{Togo:2015}
Togo,~A.; Tanaka,~I. First Principles Phonon Calculations in Materials Science.
  \emph{Scr. Mater.} \textbf{2015}, \emph{108}, 1--5\relax
\mciteBstWouldAddEndPuncttrue
\mciteSetBstMidEndSepPunct{\mcitedefaultmidpunct}
{\mcitedefaultendpunct}{\mcitedefaultseppunct}\relax
\EndOfBibitem
\bibitem[Giannozzi \latin{et~al.}(2009)Giannozzi, Baroni, Bonini, Calandra,
  Car, Cavazzoni, Ceresoli, Chiarotti, Cococcioni, Dabo, \latin{et~al.}
  others]{Giannozzi:2009}
Giannozzi,~P.; Baroni,~S.; Bonini,~N.; Calandra,~M.; Car,~R.; Cavazzoni,~C.;
  Ceresoli,~D.; Chiarotti,~G.~L.; Cococcioni,~M.; Dabo,~I. \latin{et~al.}
  QUANTUM ESPRESSO: A Modular and Open--source Software Project for Quantum
  Simulations of Materials. \emph{J. Phys.: Condens. Matter} \textbf{2009},
  \emph{21}, 395502\relax
\mciteBstWouldAddEndPuncttrue
\mciteSetBstMidEndSepPunct{\mcitedefaultmidpunct}
{\mcitedefaultendpunct}{\mcitedefaultseppunct}\relax
\EndOfBibitem
\bibitem[Dal~Corso(2014)]{DalCorso:2014}
Dal~Corso,~A. Pseudopotentials Periodic Table: From H to Pu. \emph{Comput.
  Mater. Sci.} \textbf{2014}, \emph{95}, 337--350\relax
\mciteBstWouldAddEndPuncttrue
\mciteSetBstMidEndSepPunct{\mcitedefaultmidpunct}
{\mcitedefaultendpunct}{\mcitedefaultseppunct}\relax
\EndOfBibitem
\bibitem[Troullier and Martins(1991)Troullier, and Martins]{Troullier:1991}
Troullier,~N.; Martins,~J.~L. Efficient Pseudopotentials for Plane--wave
  Calculations. \emph{Phys. Rev. B} \textbf{1991}, \emph{43}, 1993--2006\relax
\mciteBstWouldAddEndPuncttrue
\mciteSetBstMidEndSepPunct{\mcitedefaultmidpunct}
{\mcitedefaultendpunct}{\mcitedefaultseppunct}\relax
\EndOfBibitem
\bibitem[Methfessel and Paxton(1989)Methfessel, and Paxton]{Methfessel:1989}
Methfessel,~M.; Paxton,~A.~T. High--precision Sampling for Brillouin--zone
  Integration in Metals. \emph{Phys. Rev. B} \textbf{1989}, \emph{40},
  3616--3621\relax
\mciteBstWouldAddEndPuncttrue
\mciteSetBstMidEndSepPunct{\mcitedefaultmidpunct}
{\mcitedefaultendpunct}{\mcitedefaultseppunct}\relax
\EndOfBibitem
\bibitem[Allen and Dynes(1975)Allen, and Dynes]{Allen:1975}
Allen,~P.~B.; Dynes,~R.~C. Transition Temperature of Strong--coupled
  Superconductors Reanalyzed. \emph{Phys. Rev. B} \textbf{1975}, \emph{12},
  905--922\relax
\mciteBstWouldAddEndPuncttrue
\mciteSetBstMidEndSepPunct{\mcitedefaultmidpunct}
{\mcitedefaultendpunct}{\mcitedefaultseppunct}\relax
\EndOfBibitem
\bibitem[Eliashberg(1960)]{Eliashberg:1960}
Eliashberg,~G.~M. Interactions Between Electrons and Lattice Vibrations in a
  Superconductor. \emph{Sov. Phys. JETP} \textbf{1960}, \emph{11},
  696--702\relax
\mciteBstWouldAddEndPuncttrue
\mciteSetBstMidEndSepPunct{\mcitedefaultmidpunct}
{\mcitedefaultendpunct}{\mcitedefaultseppunct}\relax
\EndOfBibitem
\bibitem[Jamieson(1963)]{Jamieson:1963}
Jamieson,~J.~C. Crystal Structures Adopted by Black Phosphorus at High
  Pressures. \emph{Science} \textbf{1963}, \emph{139}, 1291--1292\relax
\mciteBstWouldAddEndPuncttrue
\mciteSetBstMidEndSepPunct{\mcitedefaultmidpunct}
{\mcitedefaultendpunct}{\mcitedefaultseppunct}\relax
\EndOfBibitem
\bibitem[Kikegawa and Iwasaki(1983)Kikegawa, and Iwasaki]{Kikegawa:1983}
Kikegawa,~T.; Iwasaki,~H. An X--ray Diffraction Study of Lattice Compression
  and Phase Transition of Crystalline Phosphorus. \emph{Acta. Cryst.}
  \textbf{1983}, \emph{B39}, 158--164\relax
\mciteBstWouldAddEndPuncttrue
\mciteSetBstMidEndSepPunct{\mcitedefaultmidpunct}
{\mcitedefaultendpunct}{\mcitedefaultseppunct}\relax
\EndOfBibitem
\bibitem[Akahama \latin{et~al.}(1999)Akahama, Kobayashi, and
  Kawamura]{Akahama:1999}
Akahama,~Y.; Kobayashi,~M.; Kawamura,~H. Simple-cubic--simple-hexagonal
  Transition in Phosphorus Under Pressure. \emph{Phys. Rev. B} \textbf{1999},
  \emph{59}, 8520--8525\relax
\mciteBstWouldAddEndPuncttrue
\mciteSetBstMidEndSepPunct{\mcitedefaultmidpunct}
{\mcitedefaultendpunct}{\mcitedefaultseppunct}\relax
\EndOfBibitem
\bibitem[Pickard and Needs(2007)Pickard, and Needs]{Pickard:2007s}
Pickard,~C.~J.; Needs,~R.~J. Structure of Phase III of Solid Hydrogen.
  \emph{Nat. Phys.} \textbf{2007}, \emph{3}, 473--476\relax
\mciteBstWouldAddEndPuncttrue
\mciteSetBstMidEndSepPunct{\mcitedefaultmidpunct}
{\mcitedefaultendpunct}{\mcitedefaultseppunct}\relax
\EndOfBibitem
\bibitem[Luo \latin{et~al.}(1993)Luo, Greene, and Ruoff]{Luo:1993}
Luo,~H.; Greene,~R.~G.; Ruoff,~A.~L. $\beta$-Po Phase of Sulfur at 162~GPa:
  X--ray Diffraction Study to 212~GPa. \emph{Phys. Rev. Lett.} \textbf{1993},
  \emph{71}, 2943--2946\relax
\mciteBstWouldAddEndPuncttrue
\mciteSetBstMidEndSepPunct{\mcitedefaultmidpunct}
{\mcitedefaultendpunct}{\mcitedefaultseppunct}\relax
\EndOfBibitem
\bibitem[Zakharov and Cohen(1995)Zakharov, and Cohen]{Zakharov:1995}
Zakharov,~O.; Cohen,~M.~L. Theory of Structural, Electronic, Vibrational, and
  Superconducting Properties of High--pressure Phases of Sulfur. \emph{Phys.
  Rev. B} \textbf{1995}, \emph{52}, 12572--12578\relax
\mciteBstWouldAddEndPuncttrue
\mciteSetBstMidEndSepPunct{\mcitedefaultmidpunct}
{\mcitedefaultendpunct}{\mcitedefaultseppunct}\relax
\EndOfBibitem
\bibitem[Gavryushkin \latin{et~al.}(2017)Gavryushkin, Litasov, Dobrosmislov,
  and Popov]{Gavryushkin:2017a}
Gavryushkin,~P.~N.; Litasov,~K.~D.; Dobrosmislov,~S.~S.; Popov,~Z.~I.
  High--pressure Phases of Sulfur: Topological Analysis and Crystal Structure
  Prediction. \emph{Phys. Status Solidi B} \textbf{2017}, \emph{254},
  1600857\relax
\mciteBstWouldAddEndPuncttrue
\mciteSetBstMidEndSepPunct{\mcitedefaultmidpunct}
{\mcitedefaultendpunct}{\mcitedefaultseppunct}\relax
\EndOfBibitem
\bibitem[Kokail \latin{et~al.}(2016)Kokail, Heil, and Boeri]{Kokail:2016}
Kokail,~C.; Heil,~C.; Boeri,~L. Search for High--$T_c$ Conventional
  Superconductivity at Megabar Pressures in the Lithium-sulfur System.
  \emph{Phys. Rev. B} \textbf{2016}, \emph{94}, 060502(R)\relax
\mciteBstWouldAddEndPuncttrue
\mciteSetBstMidEndSepPunct{\mcitedefaultmidpunct}
{\mcitedefaultendpunct}{\mcitedefaultseppunct}\relax
\EndOfBibitem
\bibitem[Liu \latin{et~al.}(2019)Liu, Wang, Chen, Lv, Sun, and Duan]{Liu:2019}
Liu,~Y.; Wang,~C.; Chen,~X.; Lv,~P.; Sun,~H.; Duan,~D. Pressure--induced
  Structures and Properties in P--S compounds. \emph{Solid State Commun.}
  \textbf{2019}, \emph{293}, 6--10\relax
\mciteBstWouldAddEndPuncttrue
\mciteSetBstMidEndSepPunct{\mcitedefaultmidpunct}
{\mcitedefaultendpunct}{\mcitedefaultseppunct}\relax
\EndOfBibitem
\bibitem[Li \latin{et~al.}(2016)Li, Wang, Liu, Zhang, Hao, Pickard, Nelson,
  Needs, Li, Huang, Errea, Calandra, Mauri, and Ma]{Li:2016-S}
Li,~Y.; Wang,~L.; Liu,~H.; Zhang,~Y.; Hao,~J.; Pickard,~C.~J.; Nelson,~J.~R.;
  Needs,~R.~J.; Li,~W.; Huang,~Y. \latin{et~al.}  Dissociation Products and
  Structures of Solid {H$_2$S} at Strong Compression. \emph{Phys. Rev. B}
  \textbf{2016}, \emph{93}, 020103(R)\relax
\mciteBstWouldAddEndPuncttrue
\mciteSetBstMidEndSepPunct{\mcitedefaultmidpunct}
{\mcitedefaultendpunct}{\mcitedefaultseppunct}\relax
\EndOfBibitem
\bibitem[Kruglov \latin{et~al.}(2017)Kruglov, Akashi, Yoshikawa, Oganov, and
  Esfahani]{Kruglov:2017}
Kruglov,~I.; Akashi,~R.; Yoshikawa,~S.; Oganov,~A.~R.; Esfahani,~M. M.~D.
  Refined Phase Diagram of the HS System with High--$T_c$ Superconductivity.
  \emph{Phys. Rev. B} \textbf{2017}, \emph{96}, 220101(R)\relax
\mciteBstWouldAddEndPuncttrue
\mciteSetBstMidEndSepPunct{\mcitedefaultmidpunct}
{\mcitedefaultendpunct}{\mcitedefaultseppunct}\relax
\EndOfBibitem
\bibitem[Errea \latin{et~al.}(2015)Errea, Calandra, Pickard, Nelson, Needs, Li,
  Liu, Zhang, Ma, and Mauri]{Errea:2015}
Errea,~I.; Calandra,~M.; Pickard,~C.~J.; Nelson,~J.; Needs,~R.~J.; Li,~Y.;
  Liu,~H.; Zhang,~Y.; Ma,~Y.; Mauri,~F. High--pressure Hydrogen Sulfide from
  First Principles: A Strongly Anharmonic Phonon-mediated Superconductor.
  \emph{Phys. Rev. Lett.} \textbf{2015}, \emph{114}, 157004\relax
\mciteBstWouldAddEndPuncttrue
\mciteSetBstMidEndSepPunct{\mcitedefaultmidpunct}
{\mcitedefaultendpunct}{\mcitedefaultseppunct}\relax
\EndOfBibitem
\bibitem[Goncharov \latin{et~al.}(2016)Goncharov, Lobanov, Kruglov, Zhao, Chen,
  Oganov, Kon{\^o}pkov{\'a}, and Prakapenka]{Goncharov:2016}
Goncharov,~A.~F.; Lobanov,~S.~S.; Kruglov,~I.; Zhao,~X.-M.; Chen,~X.-J.;
  Oganov,~A.~R.; Kon{\^o}pkov{\'a},~Z.; Prakapenka,~V.~B. Hydrogen Sulfide at
  High Pressure: Change in Stoichiometry. \emph{Phys. Rev. B} \textbf{2016},
  \emph{93}, 174105\relax
\mciteBstWouldAddEndPuncttrue
\mciteSetBstMidEndSepPunct{\mcitedefaultmidpunct}
{\mcitedefaultendpunct}{\mcitedefaultseppunct}\relax
\EndOfBibitem
\bibitem[Mishra \latin{et~al.}(2018)Mishra, Muramatsu, Liu, Geballe,
  Somayazulu, Ahart, Baldini, Meng, Zurek, and Hemley]{Zurek:2018b}
Mishra,~A.~K.; Muramatsu,~T.; Liu,~H.; Geballe,~Z.~M.; Somayazulu,~M.;
  Ahart,~M.; Baldini,~M.; Meng,~Y.; Zurek,~E.; Hemley,~R.~J. New Calcium
  Hydrides with Mixed Atomic and Molecular Hydrogen. \emph{J. Phys. Chem. C}
  \textbf{2018}, \emph{122}, 19370--19378\relax
\mciteBstWouldAddEndPuncttrue
\mciteSetBstMidEndSepPunct{\mcitedefaultmidpunct}
{\mcitedefaultendpunct}{\mcitedefaultseppunct}\relax
\EndOfBibitem
\bibitem[Bi \latin{et~al.}(2017)Bi, Miller, Shamp, and Zurek]{Zurek:2017c}
Bi,~T.; Miller,~D.~P.; Shamp,~A.; Zurek,~E. Superconducting Phases of
  Phosphorus Hydride Under Pressure: Stabilization via Mobile Molecular
  Hydrogen. \emph{Angew. Chem. Int. Ed.} \textbf{2017}, \emph{56},
  10192--10195\relax
\mciteBstWouldAddEndPuncttrue
\mciteSetBstMidEndSepPunct{\mcitedefaultmidpunct}
{\mcitedefaultendpunct}{\mcitedefaultseppunct}\relax
\EndOfBibitem
\bibitem[Shamp \latin{et~al.}(2016)Shamp, Terpstra, Bi, Falls, Avery, and
  Zurek]{Shamp:2016}
Shamp,~A.; Terpstra,~T.; Bi,~T.; Falls,~Z.; Avery,~P.; Zurek,~E. Decomposition
  Products of Phosphine Under Pressure: PH$_2$ Stable and Superconducting?
  \emph{J. Am. Chem. Soc.} \textbf{2016}, \emph{138}, 1884--1892\relax
\mciteBstWouldAddEndPuncttrue
\mciteSetBstMidEndSepPunct{\mcitedefaultmidpunct}
{\mcitedefaultendpunct}{\mcitedefaultseppunct}\relax
\EndOfBibitem
\bibitem[Flores-Livas \latin{et~al.}(2016)Flores-Livas, Amsler, Heil, Sanna,
  Boeri, Profeta, Wolverton, Goedecker, and Gross]{Flores:2016}
Flores-Livas,~J.~A.; Amsler,~M.; Heil,~C.; Sanna,~A.; Boeri,~L.; Profeta,~G.;
  Wolverton,~C.; Goedecker,~S.; Gross,~E. K.~U. Superconductivity in Metastable
  Phases of Phosphorus-hydride Compounds Under High Pressure. \emph{Phys. Rev.
  B} \textbf{2016}, \emph{93}, 020508(R)\relax
\mciteBstWouldAddEndPuncttrue
\mciteSetBstMidEndSepPunct{\mcitedefaultmidpunct}
{\mcitedefaultendpunct}{\mcitedefaultseppunct}\relax
\EndOfBibitem
\bibitem[Drozdov \latin{et~al.}(2015)Drozdov, Eremets, and
  Troyan]{Drozdov:2015b}
Drozdov,~A.~P.; Eremets,~M.~I.; Troyan,~I.~A. Superconductivity above 100 K in
  PH$_3$ at High Pressures. \emph{arXiv preprint} \textbf{2015},
  arXiv:1508.06224\relax
\mciteBstWouldAddEndPuncttrue
\mciteSetBstMidEndSepPunct{\mcitedefaultmidpunct}
{\mcitedefaultendpunct}{\mcitedefaultseppunct}\relax
\EndOfBibitem
\bibitem[Ceppatelli \latin{et~al.}(2020)Ceppatelli, Demetrio, Serrano-Ruiz,
  Dziubek, Garbarino, Jacobs, Mezouar, Bini, and Peruzzini]{Ceppatelli}
Ceppatelli,~M.; Demetrio,~S.; Serrano-Ruiz,~M.; Dziubek,~K.; Garbarino,~G.;
  Jacobs,~J.; Mezouar,~M.; Bini,~R.; Peruzzini,~M. High Pressure Synthesis of
  Phosphine from the Elements and the Discovery of the Missing
  (PH$_3$)$_2$H$_2$ Tile. \emph{Nat. Commun.} \textbf{2020}, \emph{11},
  6125\relax
\mciteBstWouldAddEndPuncttrue
\mciteSetBstMidEndSepPunct{\mcitedefaultmidpunct}
{\mcitedefaultendpunct}{\mcitedefaultseppunct}\relax
\EndOfBibitem
\bibitem[Yao and Tse(2017)Yao, and Tse]{Yao-S-review:2018}
Yao,~Y.; Tse,~J.~S. Superconducting Hydrogen Sulfide. \emph{Chem. Eur. J}
  \textbf{2017}, \emph{24}, 1769--1778\relax
\mciteBstWouldAddEndPuncttrue
\mciteSetBstMidEndSepPunct{\mcitedefaultmidpunct}
{\mcitedefaultendpunct}{\mcitedefaultseppunct}\relax
\EndOfBibitem
\bibitem[Verma and Modak(2018)Verma, and Modak]{Verma:2018a}
Verma,~A.~K.; Modak,~P. A Unique Metallic Phase of {H$_3$S} at High--pressure:
  Sulfur in Three Different Local Environments. \emph{Phys. Chem. Chem. Phys.}
  \textbf{2018}, \emph{20}, 26344--26350\relax
\mciteBstWouldAddEndPuncttrue
\mciteSetBstMidEndSepPunct{\mcitedefaultmidpunct}
{\mcitedefaultendpunct}{\mcitedefaultseppunct}\relax
\EndOfBibitem
\bibitem[Li \latin{et~al.}(2014)Li, Hao, Liu, Li, and Ma]{Li:2014}
Li,~Y.; Hao,~J.; Liu,~H.; Li,~Y.; Ma,~Y. The Metallization and
  Superconductivity of Dense Hydrogen Sulfide. \emph{J. Chem. Phys.}
  \textbf{2014}, \emph{140}, 174712\relax
\mciteBstWouldAddEndPuncttrue
\mciteSetBstMidEndSepPunct{\mcitedefaultmidpunct}
{\mcitedefaultendpunct}{\mcitedefaultseppunct}\relax
\EndOfBibitem
\bibitem[Akashi \latin{et~al.}(2015)Akashi, Kawamura, Tsuneyuki, Nomura, and
  Arita]{Akashi:2015-S}
Akashi,~R.; Kawamura,~M.; Tsuneyuki,~S.; Nomura,~Y.; Arita,~R.
  First--principles Study of the Pressure and Crystal--structure Dependences of
  the Superconducting Transition Temperature in Compressed Sulfur Hydrides.
  \emph{Phys. Rev. B} \textbf{2015}, \emph{91}, 224513\relax
\mciteBstWouldAddEndPuncttrue
\mciteSetBstMidEndSepPunct{\mcitedefaultmidpunct}
{\mcitedefaultendpunct}{\mcitedefaultseppunct}\relax
\EndOfBibitem
\bibitem[Ishikawa \latin{et~al.}(2016)Ishikawa, Nakanishi, Shimizu,
  Katayama-Yoshida, Oda, and Suzuki]{Ishikawa:2016}
Ishikawa,~T.; Nakanishi,~A.; Shimizu,~K.; Katayama-Yoshida,~H.; Oda,~T.;
  Suzuki,~N. Superconducting {H$_5$S$_2$} Phase in Sulfur--hydrogen System
  Under High--pressure. \emph{Sci. Rep.} \textbf{2016}, \emph{6}, 23160\relax
\mciteBstWouldAddEndPuncttrue
\mciteSetBstMidEndSepPunct{\mcitedefaultmidpunct}
{\mcitedefaultendpunct}{\mcitedefaultseppunct}\relax
\EndOfBibitem
\bibitem[Akashi \latin{et~al.}(2016)Akashi, Sano, Arita, and
  Tsuneyuki]{Akashi:2016a}
Akashi,~R.; Sano,~W.; Arita,~R.; Tsuneyuki,~S. Possible ``Magn{\'e}li'' Phases
  and Self-alloying in the Superconducting Sulfur Hydride. \emph{Phys. Rev.
  Lett.} \textbf{2016}, \emph{117}, 075503\relax
\mciteBstWouldAddEndPuncttrue
\mciteSetBstMidEndSepPunct{\mcitedefaultmidpunct}
{\mcitedefaultendpunct}{\mcitedefaultseppunct}\relax
\EndOfBibitem
\bibitem[Gordon \latin{et~al.}(2016)Gordon, Xu, Xiang, Bussmann-Holder, Kremer,
  Simon, K{\"o}hler, and Whangbo]{Gordon:2016}
Gordon,~E.~E.; Xu,~K.; Xiang,~H.; Bussmann-Holder,~A.; Kremer,~R.~K.;
  Simon,~A.; K{\"o}hler,~J.; Whangbo,~M.-H. Structure and Composition of the
  200~K--Superconducting Phase of {H$_2$S} at Ultrahigh Pressure: The
  Perovskite {(SH$^-$)(H$_3$S$^+$)}. \emph{Angew. Chem. Int. Ed.}
  \textbf{2016}, \emph{55}, 3682--3684\relax
\mciteBstWouldAddEndPuncttrue
\mciteSetBstMidEndSepPunct{\mcitedefaultmidpunct}
{\mcitedefaultendpunct}{\mcitedefaultseppunct}\relax
\EndOfBibitem
\bibitem[Majumdar \latin{et~al.}(2017)Majumdar, Tse, and Yao]{Majumdar:S-2017}
Majumdar,~A.; Tse,~J.~S.; Yao,~Y. Modulated Structure Calculated for
  Superconducting Hydrogen Sulfide. \emph{Angew. Chem. Int. Ed.} \textbf{2017},
  \emph{129}, 11548--11551\relax
\mciteBstWouldAddEndPuncttrue
\mciteSetBstMidEndSepPunct{\mcitedefaultmidpunct}
{\mcitedefaultendpunct}{\mcitedefaultseppunct}\relax
\EndOfBibitem
\bibitem[Majumdar \latin{et~al.}(2019)Majumdar, Tse, and Yao]{Majumdar:S-2019}
Majumdar,~A.; Tse,~J.~S.; Yao,~Y. Mechanism for the Structural Transformation
  to the Modulated Superconducting Phase of Compressed Hydrogen Sulfide.
  \emph{Sci. Rep.} \textbf{2019}, \emph{9}, 5023\relax
\mciteBstWouldAddEndPuncttrue
\mciteSetBstMidEndSepPunct{\mcitedefaultmidpunct}
{\mcitedefaultendpunct}{\mcitedefaultseppunct}\relax
\EndOfBibitem
\bibitem[Sun \latin{et~al.}(2016)Sun, Dacek, Ong, Hautier, Jain, Richards,
  Gamst, Persson, and Ceder]{materialsproject}
Sun,~W.; Dacek,~S.~T.; Ong,~S.~P.; Hautier,~G.; Jain,~A.; Richards,~W.~D.;
  Gamst,~A.~C.; Persson,~K.~A.; Ceder,~G. The Thermodynamic Scale of Inorganic
  Crystalline Metastability. \emph{Sci. Adv.} \textbf{2016}, \emph{2},
  e1600225\relax
\mciteBstWouldAddEndPuncttrue
\mciteSetBstMidEndSepPunct{\mcitedefaultmidpunct}
{\mcitedefaultendpunct}{\mcitedefaultseppunct}\relax
\EndOfBibitem
\bibitem[Toher \latin{et~al.}(2019)Toher, Oses, Hicks, and
  Curtarolo]{Toher:2019}
Toher,~C.; Oses,~C.; Hicks,~D.; Curtarolo,~S. Unavoidable Disorder and Entropy
  in Multicomponent Systems. \emph{npj Comput. Mater.} \textbf{2019}, \emph{5},
  69\relax
\mciteBstWouldAddEndPuncttrue
\mciteSetBstMidEndSepPunct{\mcitedefaultmidpunct}
{\mcitedefaultendpunct}{\mcitedefaultseppunct}\relax
\EndOfBibitem
\bibitem[Zur()]{Zurek:2016b}
Zurek, E. The Pressing Role of Theory in Studies of Compressed Matter. In
  $Handbook$ $of$ $Solid$ $State$ $Chemistry$, Vol. 5; Wiley-VCH Verlag GmbH \&
  Co., 2017; pp 571-605. DOI:10.1002/9783527691036.hsscvol5020\relax
\mciteBstWouldAddEndPuncttrue
\mciteSetBstMidEndSepPunct{\mcitedefaultmidpunct}
{\mcitedefaultendpunct}{\mcitedefaultseppunct}\relax
\EndOfBibitem
\bibitem[Li \latin{et~al.}(2017)Li, Baldini, Wang, Chen, Xu, Vermilyea, Crespi,
  Hoffmann, Molaison, Tulk, Guthrie, Sinogeikin, and Badding]{Badding:2017a}
Li,~X.; Baldini,~M.; Wang,~T.; Chen,~B.; Xu,~E.~S.; Vermilyea,~B.;
  Crespi,~V.~H.; Hoffmann,~R.; Molaison,~J.~J.; Tulk,~C.~A. \latin{et~al.}
  Mechanochemical Synthesis of Carbon Nanothread Single Crystals. \emph{J. Am.
  Chem. Soc.} \textbf{2017}, \emph{139}, 16343--16349\relax
\mciteBstWouldAddEndPuncttrue
\mciteSetBstMidEndSepPunct{\mcitedefaultmidpunct}
{\mcitedefaultendpunct}{\mcitedefaultseppunct}\relax
\EndOfBibitem
\bibitem[Mardanya \latin{et~al.}(2019)Mardanya, Singh, Huang, Chang, Su, Lin,
  Agarwal, and Bansil]{Mardanya:2019}
Mardanya,~S.; Singh,~B.; Huang,~S.-M.; Chang,~T.-R.; Su,~C.; Lin,~H.;
  Agarwal,~A.; Bansil,~A. Prediction of Threefold Fermions in a Nearly Ideal
  Dirac Semimetal BaAgAs. \emph{Phys. Rev. Mater.} \textbf{2019}, \emph{3},
  071201(R)\relax
\mciteBstWouldAddEndPuncttrue
\mciteSetBstMidEndSepPunct{\mcitedefaultmidpunct}
{\mcitedefaultendpunct}{\mcitedefaultseppunct}\relax
\EndOfBibitem
\bibitem[Minkov \latin{et~al.}(2020)Minkov, Prakapenka, Greenberg, and
  Eremets]{Minkov:2020a}
Minkov,~V.~S.; Prakapenka,~V.~B.; Greenberg,~E.; Eremets,~M.~I. A Boosted
  Critical Temperature of 166 K in Superconducting D$_3$S Synthesized from
  Elemental Sulfur and Hydrogen. \emph{Angew. Chem. Int. Ed.} \textbf{2020},
  \emph{59}, 18970--18974\relax
\mciteBstWouldAddEndPuncttrue
\mciteSetBstMidEndSepPunct{\mcitedefaultmidpunct}
{\mcitedefaultendpunct}{\mcitedefaultseppunct}\relax
\EndOfBibitem
\bibitem[ccr()]{ccr}
http://hdl.handle.net/10477/79221\relax
\mciteBstWouldAddEndPuncttrue
\mciteSetBstMidEndSepPunct{\mcitedefaultmidpunct}
{\mcitedefaultendpunct}{\mcitedefaultseppunct}\relax
\EndOfBibitem
\end{mcitethebibliography}
\end{document}